\documentclass[journal,draftclsnofoot,onecolumn,12pt]{IEEEtran}
\usepackage{amsthm,amssymb,graphicx,multirow,amsmath,color,amsfonts}
\usepackage[update,prepend]{epstopdf}
\usepackage[noadjust]{cite}
\usepackage[latin1]{inputenc}
\usepackage{tikz}
\usepackage{bbm} 
\usepackage{pdfpages}
\usepackage{multirow}
\usepackage{subfig}
\usepackage{comment}
\def\nb0{{\mathbf{0}}}
\def\nb1{{\mathbf{1}}}

\newtheorem{lemma}{Lemma}

\allowdisplaybreaks 

\usepackage{setspace}	

\begin{document}
\graphicspath{{./figures/}}
\title{
Dedicating Cellular Infrastructure for Aerial Users: Advantages and Potential Impact on Ground Users}
\author{
Lin Chen, Mustafa A. Kishk, {\em Member, IEEE}{,} and Mohamed-Slim Alouini, {\em Fellow, IEEE}
\thanks{
Lin Chen is with the Department of Information Engineering, The Chinese University of Hong Kong (CUHK), Hong Kong.
Mustafa A. Kishk is with the Department of Electronic Engineering, National University of Ireland, Maynooth, W23 F2H6, Ireland.
Mohamed-Slim Alouini is with KAUST, CEMSE division, Thuwal 23955-6900, Saudi Arabia. (e-mail: lin.chen@link.cuhk.edu.hk; mustafa.kishk@mu.ie; slim.alouini@kaust.edu.sa). 
} 
\vspace{-4mm}}

\maketitle

\begin{abstract}
\textcolor{black}{
A new generation of aerial vehicles is hopeful to be the next frontier for the transportation of people and goods, becoming even as important as ground users in the communication systems. 
To enhance the coverage of aerial users, appropriate adjustments should be made to the existing cellular networks that mainly provide services for ground users by the down-tilted antennas of the terrestrial base stations (BSs). It is promising to up-tilt the antennas of a subset of BSs for serving aerial users through the mainlobe. With this motivation, in this work, we use tools from stochastic geometry to analyze the coverage performance of the adjusted cellular network (consisting of the {up-tilted} BSs and the {down-tilted} BSs). 
Correspondingly, we present exact and approximate expressions of the signal-to-interference ratio (SIR)-based coverage probabilities for users in the sky and on the ground, respectively.
Numerical results verify the analysis accuracy and clarify the advantages of up-tilting BS antennas on the communication connectivity of aerial users without the potential adverse impact on the {quality of service (QoS)} of ground users. 
Moreover, it is unveiled that there exists an optimal value of the up-tilted/down-tilted BS density ratio for maximizing the coverage probability of the aerial or ground users.
}
 
\end{abstract}

\section{Introduction}
In recent years, aerial transportation has seen unprecedented advances since the terrestrial traffic congestion and the constraints of public transportation infrastructure. Transportation of (i) goods through unmanned aerial vehicles (UAVs) and (ii) people through flying cars is not a futuristic dream anymore~\cite{cvitanic2020drone,mozaffari2019tutorial}. 

To become a reality, it is important to provide a strong and reliable connection for all kinds of aerial transportation to ensure (i) safety and control of UAVs and (ii) coverage for mobile users in flying cars.
However, the current cellular infrastructure is primarily designed to serve users spatially distributed on the ground. In particular, the antennas of the terrestrial base stations (BSs) are {down-tilted} and completely direct towards the ground users. Consequently, the aerial users can only get sidelobe gain from the current cellular infrastructure, which might not be enough to ensure the full coverage of such a new type of users. Therefore, there is an urgent need to modify the current cellular network to meet the connectivity requirement of aerial users (e.g., UAVs for delivery or surveillance and mobile equipment held by people in flying cars).

\subsection{Related Work}\label{sec:related}
\textcolor{black}{
Aerial communication equipment-enabled communication systems have recently attracted much research interest~\cite{yi2022joint,peng2021leopard,vilor2020optimal,kishk2020aerial,matracia2021coverage,qin2020performance}. For example, by employing geographic information, the authors in~\cite{yi2022joint} jointly optimized the 3D position and power allocation of the UAV relay to improve communication capacity. In~\cite{peng2021leopard}, UAVs were served as BSs, whose placement is based on the prediction of the user equipment (UE) movement, to provide seamless communication services for the flash mobile crowds.  
Ref. \cite{vilor2020optimal} formulated an optimization problem to design the trajectory of a UAV by maximizing the minimum rate of the downlink (from a UAV to a ground UE). However, in the above works, UAVs either act as BSs or relays to enhance the quality of service (QoS) of ground users, while the coverage probability of UAVs (regarded as UEs, i.e., UAV-UEs) is not considered.}

\textcolor{black}{
The QoS of aerial communication equipment served by the existing cellular network also has some work.
The authors in~\cite{feasibility1,feasibility2} discussed the technical feasibility of leveraging the established cellular network for supporting the connectivity of UAVs in a cost-effective manner. 
Modeling the actual radiation pattern in the vertical plane of the BSs equipped with uniform linear antennas (ULAs), Ref.~\cite{lyu2019network} provided the uplink and downlink coverage performance analysis for the UAV-UE in cellular networks composed of regularly-distributed BSs.
}

\textcolor{black}
{On the other hand, adjusting the current network structure to serve aerial users has been investigated recently, e.g.,~\cite{model0,height1,height2,downtilt1,downtilt2,antenna1,antenna2}.
The authors in~\cite{height1} found an optimal value of the UAV height to ensure connectivity. 
The height-dependent path-loss exponent and the small-scale fading were further considered in~\cite{height2} to analyze the coverage probability of the aerial users. In addition to optimizing the height of the aerial users, the parameters of BSs have been modified to improve the aerial coverage probability~\cite{downtilt1,downtilt2,antenna1,antenna2}.
Ref. \cite{downtilt1,downtilt2} proposed to reduce the {down-tilt} angle and scale the beamwidth of BS antennas to allow part of BSs to provide services through mainlobes for both aerial users and ground users. However, once such BSs are associated with ground users, they unavoidably interfere with aerial users from the mainlobes and vice versa. In order to suppress interference, antenna patterns were designed in~\cite{antenna1,antenna2}.
}

\textcolor{black}{
It is worth noting that the existing literature on improving the QoS of aerial users mainly focuses on optimizing the height of BSs/UAVs and the down-tilted angle/beamwidth. However, in these solutions, the BSs tilting their antennas downward still aims at serving the ground users.
In fact, as a result of the on-growing set of applications of UAVs, it is anticipated to see a continuous increase in the number of aerial users, even becoming comparable to the number of ground users in the communication systems.
Therefore, it is reasonable to design a cellular network composed of two types of BSs for serving ground users and aerial users, respectively.
}

\subsection{Contributions}\label{sec:cont}

\textcolor{black}{
Motivated by the above discussion, in the paper, we propose to up-tilt the antennas of a fraction of BSs to ensure the coverage of aerial users. Intuitively, the {up-tilted} BSs enable aerial users to receive higher power through the mainlobes and only interfere with ground users from the sidelobes. This implies that the proposal can {\em increase the received power at aerial users} and {\em decrease the interference power at ground users}. 
However, 
this proposal also involves some new technical challenges related to characterizing the interference from both the sidelobes of down-tiled BSs and the sidelobes/mainlobes of up-tilted BSs for aerial users and the opposite for ground users. 
We use the {signal-to-interference ratio (SIR)}-based coverage probability as a performance metric to quantify the impact of the coexistence of up-tilted BSs and down-tilted BSs on aerial users and grounds users.
Employing the stochastic geometry approach, we seek a reasonable solution to improve the coverage probability of aerial users without deteriorating the {QoS} of ground users. 
The main contributions of this paper are listed as follows:
\begin{itemize}
\item 
We propose a new method to ensure the connectivity of aerial users, i.e. converting a part of the {down-tilted} BSs into {up-tilted} BSs, which {makes} it possible for the cellular network to transmit signals to aerial users through mainlobes.
\item 
Using stochastic geometry, we derive the expressions of the SIR-based coverage probabilities of aerial users and ground users to evaluate the effectiveness of the proposed cellular network. We also verify the analysis accuracy by extensive Monte Carlo simulations.
\item 
With the improvement on the coverage probabilities in numerical results compared with the current network comprising only down-tilted BSs, we show that the proposed network is effective to increase the QoS of aerial users without lowering that of ground users.
\item 
We further explore the impact of system parameters, including the up-tilted/down-tilted BS density ratio, the up-tilted angle/beamwidth, and the heights of BSs and aerial users on the coverage performance.
These analyses provide insights into the design of future networks to
achieve wide coverage for both aerial users and ground users.
\end{itemize}
}

The rest of this paper is structured as follows. We introduce the proposed cellular network, the corresponding stochastic geometry-based model, and the performance metrics in Sec.~\ref{sec:sys}. In Sec.~\ref{sec:analy}, we derive the expressions of the performance metrics. Then, we present and discuss the numerical results in Sec.~\ref{sec:simu}. Finally, we conclude our work in Sec.~\ref{sec:conclusion}. Table~\ref{tab:TableOfNotations} summarizes the notations in this paper.

\begin{table*}[ht]\caption{Table of notations}
\centering
\begin{center}
\resizebox{\textwidth}{!}{
\renewcommand{\arraystretch}{1.4}
    \begin{tabular}{ {c} | {c} }
    \hline
        \hline
    \textbf{Notation} & \textbf{Description} \\ \hline

    $h_{\rm T}$; $h_{\rm a}$; $h_{\rm g}$ & The altitude of the terrestrial BSs; the aerial users; the ground users\\ \hline
    $\Psi_{\rm T}$; $\Psi_{\rm a}$; $\Psi_{\rm g}$ & The PPP modeling the locations of the terrestrial BSs; the aerial users; the ground users \\ \hline
    $\lambda_{\rm T}$; $\lambda_{\rm a}$; $\lambda_{\rm g}$ & The density of $\Psi_{\rm T}$; $\Psi_{\rm a}$; $\Psi_{\rm g}$  \\ \hline
    $G_{\rm M}$; $G_{\rm S}$ & Mainlobe gain; sidelobe gain provided by BSs\\ \hline 
    $\Psi_{\rm U}$; $\Psi_{\rm D}$ & The PPP modeling the locations of BSs with the {up-tilted} antennas; the {down-tilted} antennas \\ \hline   
    $\theta_{\rm U}$; $\theta_{\rm D}$ & The {up-tilt} angle; the {down-tilt} angle of the BS antenna\\ \hline
    $\varphi_{\rm U}$; $\varphi_{\rm D}$ & The vertical antenna beamwidth of the {up-tilted} BSs; the {down-tilted} BSs\\ \hline
    $\delta$ & The fraction of the BSs that direct their antennas towards aerial users by up-tilting.  \\ \hline
    $\lambda_{\rm U}$; $\lambda_{\rm D}$ & The density of $\Psi_{\rm U}$; $\Psi_{\rm D}$. Note that $\lambda_{\rm U}=\delta\lambda_{\rm T}$ and $\lambda_{\rm D}=(1-\delta)\lambda_{\rm T}$  \\ \hline 
        $\mathcal{P}^{\rm L}(r)$ & The probability that a link with horizontal distance $r$ is clear of any {blockage} \\ \hline
        $\mathcal{P}^{\rm N}(r)$ & The probability that a link is obstructed by at least one blockage, where $\mathcal{P}^{\rm N}(r)=1-\mathcal{P}^{\rm L}(r)$ \\ \hline
           $\rm UML$; $\rm UMN$; $\rm USL$; $\rm USN$ & The BS using {up-tilted} antenna provides mainlobe gain or sidelobe gain with LoS or NLoS transmission\\ \hline
           $\rm DML$; $\rm DMN$; $\rm DSL$; $\rm DSN$ & The BS using {down-tilted} antenna provides mainlobe gain or sidelobe gain with LoS or NLoS transmission\\ \hline
           $b$; $w$ & The type of serving BSs, where $b=b_1b_2b_3$ with $b_1\in\left\{ \rm U,\rm D \right\}$, $b_2\in\left\{ \rm M,\rm S \right\}$, and $b_3\in\left\{ \rm L,\rm N \right\}$; the type of interfering BSs\\ \hline
    \end{tabular}}
\end{center}
\label{tab:TableOfNotations}
\end{table*}

\section{System Model}\label{sec:sys}
In this section, we introduce the idea of switching a subset of the cellular infrastructure into a fully dedicated network for serving aerial users, followed by the channel model, the association policy between the BSs and the users, and the performance metrics.

\subsection{Network Model}\label{subsec:network}
\begin{figure}
\centering
\includegraphics[width=0.7\columnwidth]{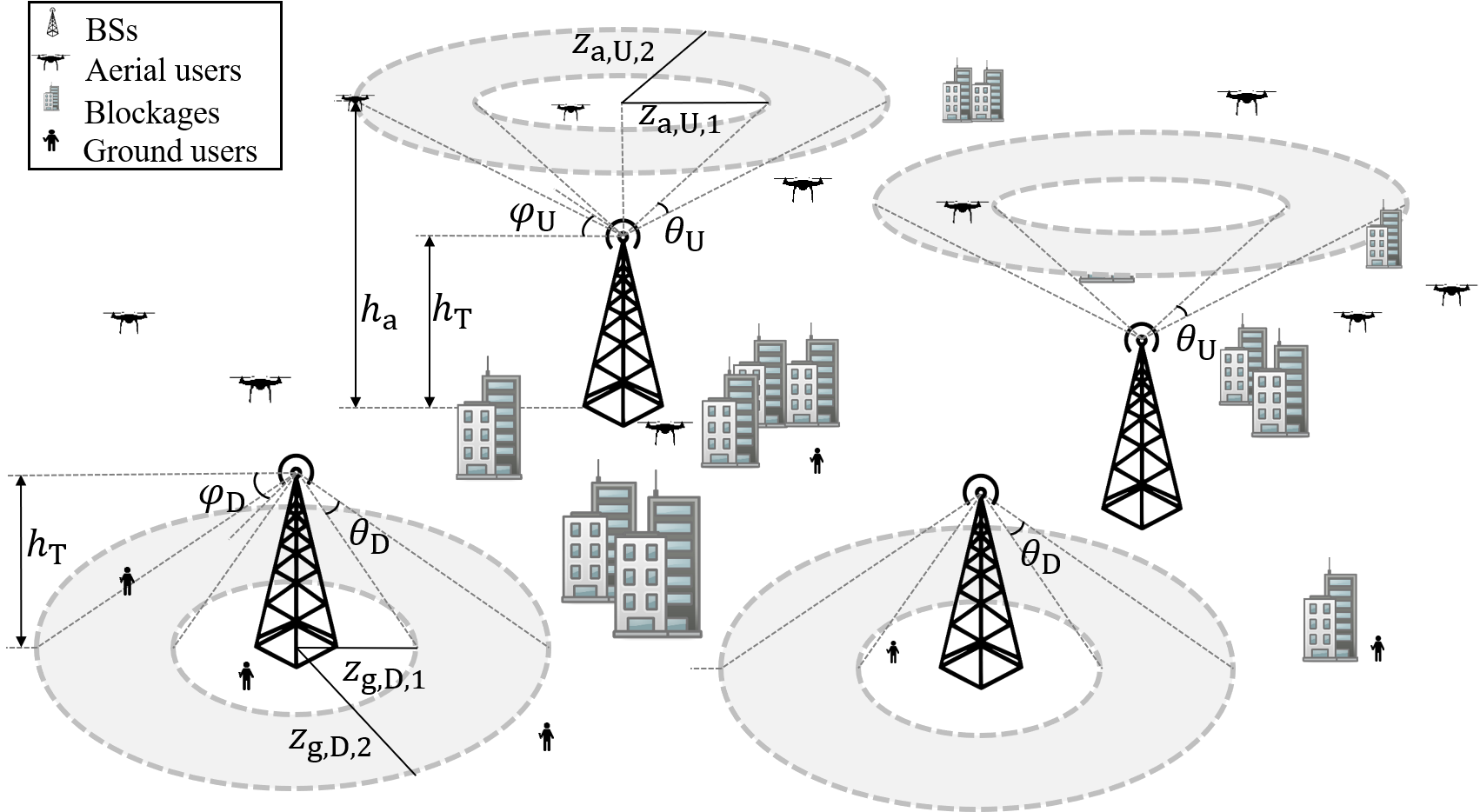}
\caption{Depiction of a fraction of the terrestrial BSs that direct their antennas towards aerial users. }
\label{fig:system}
\end{figure}

\textcolor{black}{
We consider a cellular network consisting of terrestrial BSs, aerial users, and ground users with particular altitudes $h_{\rm T}$, $h_{\rm a}$, and $h_{\rm g}$, respectively. We model their 2D locations as three independently homogeneous PPPs (HPPPs): $\Psi_{\rm T}=\{t_i\}\in\mathbb{R}^2$ with density $\lambda_{\rm T}$, $\Psi_{\rm a}=\{a_i\}\in\mathbb{R}^2$ with density $\lambda_{\rm a}$, and $\Psi_{\rm g}=\{g_i\}\in\mathbb{R}^2$ with density $\lambda_{\rm g}$, respectively. In fact, the aerial users are movable and have different altitudes. The 2D-PPP modelling of aerial users can be considered as an approximation to a scenario in which the altitudes of aerial users are uniformly distributed within a range of heights and $h_{\rm a}$ is the average altitude.\footnote{\textcolor{black}{The analysis of aerial communication devices with different altitudes matches that with the same altitude (equal to the average altitudes of aerial communication devices)~\cite{approximate,ABSheight}}.}
}
Each BS serves one single user in a time-frequency slot.
Besides, we assume that the antenna radiation patterns of BSs are omnidirectional in the horizontal plane and directional in the vertical plane, while all users employ omnidirectional antennas.
As illustrated in Fig.~\ref{fig:system}, a fraction ($\delta$) of terrestrial BSs are dedicated to serving aerial users by up-tilting their beams with angle $\theta_{\rm U}$ and vertical beamwidth $\varphi_{\rm U}$, where $0\leq\delta\leq1$. The density of the BSs with an {up-tilt} angle is denoted by $\lambda_{\rm U}=\delta\lambda_{\rm T}$.
The rest of BSs have {down-tilt} angle $\theta_{\rm D}$ and vertical beamwidth $\varphi_{\rm D}$ with density $\lambda_{\rm D}=(1-\delta)\lambda_{\rm T}$.

To this end, our main purpose throughout this paper is to analyze the considered setup in terms of four specific parameters:
\begin{itemize}
\item The fraction of cellular BSs that should direct their antennas towards aerial users: $\delta$.
\item The antenna pattern of BSs, i.e. {up-tilt} or {down-tilt} angles of the BSs and their vertical beamwidths: $\theta_{\rm U}$, $\theta_{\rm D}$, $\varphi_{\rm U}$, and $\varphi_{\rm D}$.
\item The heights of elements in the cellular network: $h_{\rm T}$, $h_{\rm a}$, and $h_{\rm g}$.
\item The probability that the {SIR at} the typical user is {above} a predefined threshold: $\mathcal{P}^{\rm cov}$.
\end{itemize}
Given the values of the mainlobe and sidelobe gain of the BSs, we aim to derive the expression of $\mathcal{P}^{\rm cov}$ as a function of $\delta$, $\theta_{\rm U}$, $\theta_{\rm D}$, $\varphi_{\rm U}$, $\varphi_{\rm D}$, $h_{\rm T}$, $h_{\rm a}$, and $h_{\rm g}$ for aerial users and ground users, so as to study how the ground mobile users can be affected when taking a fraction of their cellular infrastructure to support aerial communications. The expression offers some instructive information for designing the cellular network in different kinds of communication environments. 

\textcolor{black}{Without loss of generality, the following analysis is for a typical user above or at the origin (i.e., the typical aerial-user and the typical ground-user)~\cite{haenggi2012stochastic}. The distance from a BS, e.g., $t_i\in \Psi_{\rm T}$, to the typical user, i.e., $r_i=||t_i||$, refers to the horizontal distance, unless otherwise stated.
}


\subsection{Channel Model}\label{subsec:channel}
In this subsection, we present the antenna gain of the up-tilted BSs and down-tilted BSs, respectively. We also consider the characteristics of line-of-sight (LoS) transmission and non LoS (NLoS) transmission for channels from the terrestrial BSs to aerial users (T2A) and to ground users (T2G) when calculating the path loss. Furthermore, we use the Nakagami-$m$ fading model to describe the small-scale fading. Then, we provide the received power and the SIR of {a typical user} in the sky and on the ground.

Based on the aforementioned antenna angles and beamwidths of BSs, each user experiences either mainlobe gain or sidelobe gain from an up-tilted/down-tilted BS, which depends on their positions. 
For illustration, we denote the type of users by $v\in\{\rm a,g\}$, where $v={\rm a}$ represents the aerial users and $v={\rm g}$ represents the ground users.
The antenna gain provided by a {up-tilted} BS with horizontal distance $r$ to a $v$-type user provides antenna gain is given by
\begin{equation}
\label{eq:gainup}
G_{v,\rm U}(r)= 
\begin{cases}
G_{\rm M} 
 & \text{\rm if} ~  z_{v,{\rm U},1}<r\le z_{v,{\rm U},2} \\
G_{\rm S} & \text{\rm otherwise},
\end{cases}
\quad
v\in \left \{ \rm a,g \right \},
\end{equation}
where $z_{v,{\rm U},1}=\min  \left \{ 0,\left ( h_v-h_{\rm T}  \right )\cot(\theta_{\rm U}+\frac{\varphi_{\rm U}}{2} ) \right \}$ and $z_{v,{\rm U},2}=\min  \left \{ 0,\left ( h_v-h_{\rm T}  \right )\cot(\theta_{\rm U}-\frac{\varphi_{\rm U}}{2} ) \right \} $.
As depicted in Fig.~\ref{fig:system}, $(z_{v,{\rm U},1},z_{v,{\rm U},2}]$ {is the range of the mainlobe coverage area defined by a up-tilted BS on a horizontal plane at a specific height ($h_{\rm a}$ or $h_{\rm g}$)}.
Similarly, the antenna gain provided by a down-tilted BS with distance $r$ to the typical user at origin ($v={\rm g}$) or above origin ($v={\rm a}$) is given by 
\begin{equation}
\label{eq:gaindown}
G_{v,\rm D}(r)= 
\begin{cases}
G_{\rm M}
 & \text{\rm if} ~  z_{v,{\rm D},1}<r\le z_{v,{\rm D},2}\\
G_{\rm S} & \text{\rm otherwise},
\end{cases}
\quad v\in \left \{ \rm a,g \right \},
\end{equation}
where $z_{v,{\rm D},1}=\min  \left \{ 0,\left ( h_{\rm T}-h_v  \right )\cot(\theta_{\rm D}+\frac{\varphi_{\rm D}}{2} ) \right \} $ and $z_{v,{\rm D},2}=\min  \left \{ 0,\left ( h_{\rm T}-h_v  \right )\cot(\theta_{\rm D}-\frac{\varphi_{\rm D}}{2} ) \right \} $. $(z_{v,{\rm D},1}, z_{v,{\rm D},2}]$ {describes the mainlobe coverage area of a down-tilted BS} in Fig.~\ref{fig:system}.
Due to the fact that the BSs with {up-tilted} antennas are fully dedicated to serving aerial users, it is reasonable that $\theta_{\rm U}-\frac{\varphi_{\rm U}}{2} >0$ and $\theta_{\rm D}-\frac{\varphi_{\rm D}}{2} >0$ in (\ref{eq:gainup}) and (\ref{eq:gaindown}). 
Thus, $z_{{\rm g,U},1}=z_{{\rm g,U},2}=z_{{\rm a,D},1}=z_{{\rm a,D},2}=0$,
which means the {up-tilted} BSs can provide either mainlobe or sidelobe gain for aerial users while only sidelobe gain for ground users. Likewise, the aerial users can not be served through mainlobes of the {down-tilted} BSs. Hence, (\ref{eq:gainup}) and (\ref{eq:gaindown}) can be simplified as
\begin{equation}\label{eq:GS}
\begin{split}
G_{\rm g,U}(r)=G_{\rm a,D}(r)=G_{\rm S}.
\end{split}
\end{equation}

\begin{figure}%
 \centering
 \subfloat[]{\includegraphics[width=0.55\columnwidth]{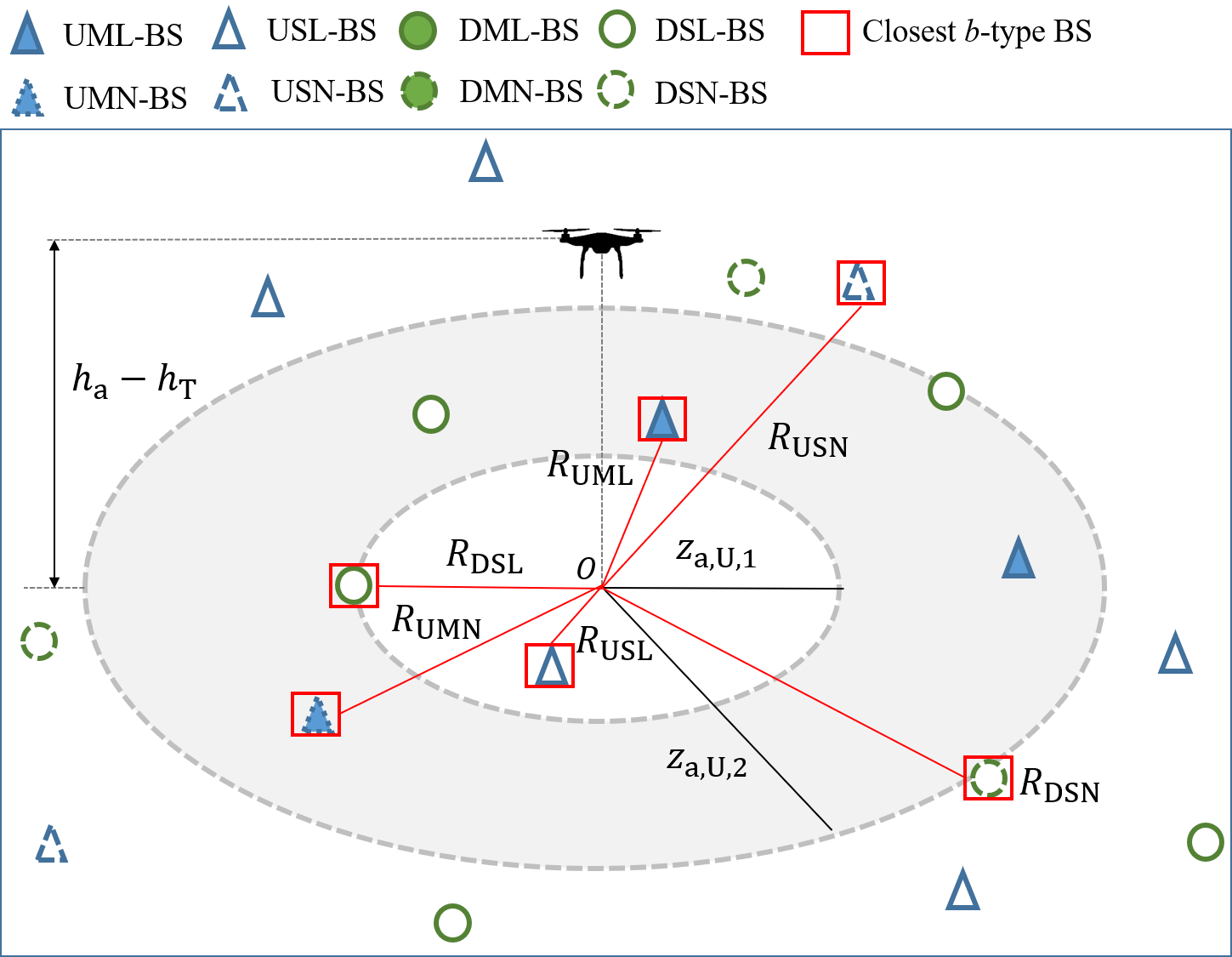}\label{fig:bs1}}\\
 \subfloat[]{\includegraphics[width=0.55\columnwidth]{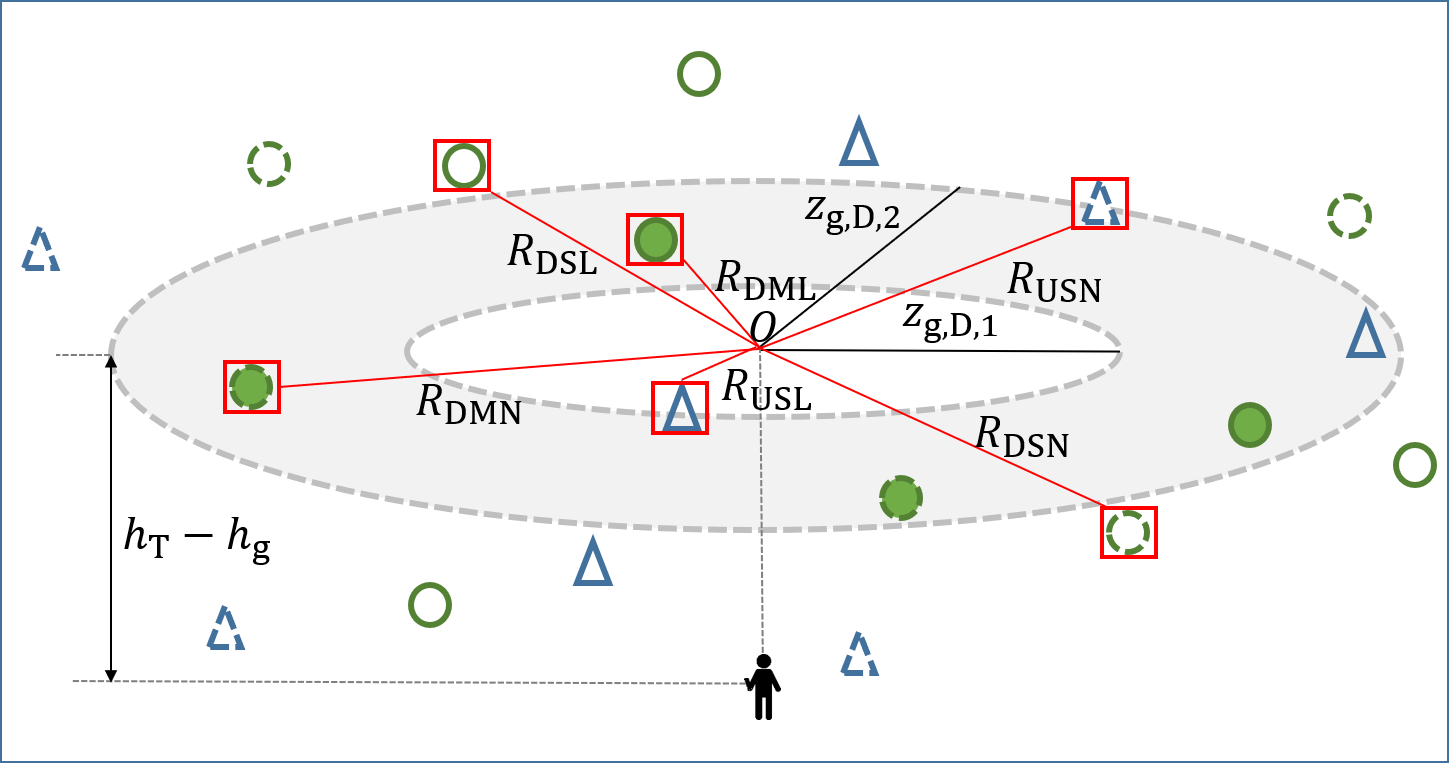}\label{fig:bs2}}
 \caption{Types of BSs from the perspective of (a) aerial users {and} (b) ground users. }%
 \label{fig:bs} 
\end{figure}

The blockages in the communication environment lead to the LoS and NLoS links. We consider the LoS and NLoS conditions for both the T2A channel and the T2G channel, whose occurrence probabilities depend on the environment and the altitudes of the transceivers. The probability of the LoS transmission, denoted by $\mathcal{P}^{\rm L}_v(r)$, is given by~\cite{ITU2013propagationLoS}
\begin{equation}\label{eq:PLoS}
\begin{split}
\mathcal{P}^{\rm L}_v(r)&=\prod_{n=0}^{\mathcal{N}} \left [ 1-\exp \left ( -\frac{\left [  h_{\rm T}-\frac{\left (n+0.5  \right  )\left ( h_{\rm T}-h_v \right )  }{\mathcal{N}+1} \right ] ^{2} }{2\gamma^{2} }  \right )  \right ] ,~
\mathcal{N}=\left \lfloor \frac{r\sqrt{\alpha \beta }}{1000} -1  \right \rfloor ,
\end{split}
\end{equation}
where $v\in \left \{ \rm a,g \right \}$ stands for the type of the user, $r$ is the horizontal distance between a BS and a user, and the properties of the environment are described by the three constants $\left \{ \alpha, \beta, \gamma \right \}$. \textcolor{black}{$\alpha$ is the ratio of the building area to the total land area, $\beta$ represents the mean number of buildings per $\rm km^2$, and the scale parameter $\gamma$ is related to the Rayleigh probability density function (PDF), i.e., $f(H) = \frac{H}{\gamma^2} \exp\left ( - \frac{H^2}{2\gamma^2}\right )$, where $H$ is the building height and $f(H)$ is the distribution of $H$.} 
Correspondingly, the probability of NLoS transmission is $\mathcal{P}^{\rm N}_v(r)=1-\mathcal{P}^{\rm L}_v(r)$.

\textcolor{black}{For simplicity, we denote
the antenna direction (up-tilt or down-tilt), the mainlobe or sidelobe gain from a BS to the typical user, and the LoS or NLoS condition of the channel between a BS and the typical user as $w_1\in\left \{ \rm U, \rm D \right \} $,  $w_2 \in \left \{ \rm M, \rm S \right \} $, and $w_3 \in \left \{ \rm L, \rm N \right \} $, respectively. Then, the BSs set (denoted by $W$) can be divided into 8 types, i.e. $W=\left \{w\right \}=\left \{ \rm UML, UMN, USL, USN, DML, DMN, DSL, DSN \right \} $, where $w=w_1w_2w_3$. Notations $b$, $b_1$, $b_2$, and $b_3$ have the same meaning as $w$, $w_1$, $w_2$, and $w_3$, respectively, while $b$ (or $w$) represents the type of the serving (or interfering) BS in the following. More details of the notations are provided in Table~\ref{tab:TableOfNotations}.
Particularly, even though we classify the BSs into 8 types, the $w$-type BSs still follow the PPP distribution due to the independent thinning property of PPP~\cite{haenggi2012stochastic}, i.e., $\Psi_{\rm T}=\Psi_{\rm U}\cup \Psi_{\rm D},
\Psi_{\rm U}=\Psi_{\rm UML}\cup \Psi_{\rm UMN}\cup \Psi_{\rm USL}\cup \Psi_{\rm USN}, \Psi_{\rm D}=\Psi_{\rm DML}\cup \Psi_{\rm DMN}\cup \Psi_{\rm DSL}\cup \Psi_{\rm DSN}$, and $\Psi_{b}\cap \Psi_{w}=\varnothing$ ($b,w\in W, b\neq w$).}
Furthermore, (\ref{eq:GS}) simplifies the BSs set into $W_{\rm a}$ for aerial users and $W_{\rm g}$ for ground users, as shown in Fig.~\ref{fig:bs}, where
\begin{equation}\label{eq:BSset}
\begin{split}
W_{\rm a}&=\left \{ \rm UML, UMN, USL, USN, DSL, DSN \right \} ,\\
W_{\rm g}&=\left \{ \rm DML, DMN, DSL, DSN, USL, USN \right \} .
\end{split}
\end{equation}

The path loss between a typical $v$-user and a BS located at $t_i$ (with horizontal distance $r_i=||t_i||$ and type $w=w_1w_2w_3$) is given by 
\begin{equation}\label{eq:lsfading}
\begin{split}
\zeta_{v,w_3}(r_i)&=\eta_{w_3}d^{{-\alpha_{w_3}}}_{v,i}=\eta_{w_3}\left [ r_i^{2}+\left ( h_v-h_{\rm T} \right )^{2}   \right ]^{{-\alpha_{w_3}/2}}, 
v\in\left \{ \rm a, g \right \}, w_3 \in \left \{ \rm L,\rm N \right \},
\end{split}
\end{equation}
where $w_3$ represents the characteristics of the link, i.e., LoS or NLoS, $\alpha_{w_3}$ is the path-loss exponent corresponding to the link characteristics $w_3$, $\eta_{w_3}$ is a constant parameter representing the path loss at the reference distance $d=1~\rm m$, $d_{v,i}$ is the Euclidean distance between the typical $v$-user and the BS $i$, where $d_{v,i}=\sqrt{r_i^{2}+\left ( h_v-h_{\rm T} \right )^{2}} $.
The independent small-scale fading denoted by $\Omega_{w_3,i}$, follows Gamma distribution with $\mathbb{E}\left \{ \Omega_{w_3,i}  \right \} =1$. We adopt the widely-used Nakagami-$m$ fading model with shaping parameters given by $m_{\rm L}$ and $m_{\rm N}$ for LoS and NLoS links, respectively. The PDF of $\omega_{w_3,i}$ is given by~\cite{Nakagami}
\begin{align}\label{eq:ssfading}
f_{\Omega_{w_3,i}}(\omega )=\frac{m_{w_3}^{m_{w_3}} \omega^{m_{w_3}-1}}{\Gamma(m_{w_3}) } e^{-m_{w_3}\omega }, w_3 \in \left \{ \rm L, \rm N \right \},
\end{align}
where {$\Gamma(m)$} is the Gamma function and {$\Gamma\left ( m \right )=\int_{0}^{\infty} t^{m-1}e^{-t}\mathrm{d}t$~\cite{davis1959leonhard}}.
The received power at the typical $v$-user from a BS in $\Psi_w$, with horizontal distance $r_i$, is given by
\begin{align}\label{eq:powerrec}
P_{v,w}^{\rm r}(r_i)=P_{v,w_1w_2w_3}^{\rm r}(r_i)=P^{\rm t}G_{w_2}\zeta_{v,w_3}\left ( r_i \right ) \Omega_{w_3,i},
\end{align}
where $P^{\rm t}$ is the constant transmission power.
Therefore, 
when the typical $v$-user is associated with a BS located at $t_0\in \Psi_{b}$ (with horizontal distance $r_0=||t_0||$), the interference power from all BSs except the serving BS, denoted by $I_{v|r_0}$, is given by 
\begin{align}\label{eq:interference}
I_{v|r_0}=\sum_{w\in W_v} I_{v,w|r_0}
=\sum_{w\in W_v}\sum_{i,t_i\in \Psi_w\setminus \left \{ t_0 \right \} }P_{v,w}^{\rm r}(r_i),
\end{align}
where $I_{v,w|r_0}$ is the interference from all interfering $w$-BSs.
Correspondingly, the instantaneous {SIR} is given by
\begin{align}\label{eq:sinrvb}
{\rm {SIR}}_{v}^{b}=\frac{P_{v,b}^{\rm r}(r_0)}{I_{v \mid r_0} } ,
\end{align}
where $v=\{\rm a,g\}$ is the type of the typical user and $b\in W_v$ is the type of the serving BS.

\subsection{Association Policy}\label{subsec:asso}

Following the discussion in Sec.~\ref{subsec:channel}, we introduce an association policy based on the average received power. In specific, the typical user is associated with the BS that provides the {\em strongest average received power}. We denote the horizontal distance between the typical user and its serving BS as $r_0$. It is worth noting that serving BS is not always the closest one in the BSs set $W$ since signals from different types of BSs experience different channel gain. For instance, a UML-BS is able to transmit the signal from mainlobe through LoS links, which compensates for the long-distance path loss and thus {may} provide stronger average power, compared with a closer USN-BS. 
By noting $\mathbb{E}\left \{ \Omega_{w_3,i}  \right \} =1$ in \eqref{eq:powerrec}, the average received power from a BS in $\Psi_w$ with distance $r_i$ to the typical $v$-user is given by 
\begin{align}\label{eq:averPrec}
\bar{P}_{v,w}^{\rm r}(r_i) = P^{\rm t}G_{w_2}\zeta_{v,w_3}\left ( r_i \right ).
\end{align}
It can be seen from \eqref{eq:averPrec} that, for the same type of BSs with the same channel characteristics, the average received power only depends on the path loss, which is a monotonically decreasing function of the propagation distance. Clearly, the closest BS in each type of BSs can provided the strongest average received power compared with the rest BSs with the same type. Therefore, the serving BS must be one of the closest BSs from each type of BSs, which are highlighted by red squares in Fig.~\ref{fig:bs}. 
Let $r_0$ denote the horizontal distance between the serving BS and the typical $v$-user. The association policy can be mathematically expressed as
\begin{equation}\label{eq:r0}
\begin{split}
\textcolor{black}{
r_0= \underset{R_b,\,b\in W_v}{\arg\; \max}\left \{ \bar{P}_{v,b}^{\rm r}(R_b) \right \},
~R_b=\min_{i,\,t_i\in \Psi_{b}}\{r_i\},
}
\end{split}
\end{equation}
where $R_b$ is the closest horizontal distance between BSs in $\Psi_b$ and the typical $v$-user and $\bar{P}_{v,b}^{\rm r}(R_b)$ is the corresponding strongest average received power from a BS in $\Psi_b$ and is given in \eqref{eq:averPrec}.

\subsection{Performance Metrics}\label{subsec:metrics}
We adopt the coverage probability, i.e., the probability that SIR is above a predefined threshold, as a metric to quantify the performance of the proposed cellular network. Let $\mathcal{C}_{v}$ denote the event that the typical $v$-user is in coverage.
Given threshold $\tau$, the coverage probability of the T2A or T2G link can be defined as follows.
\begin{equation}
\begin{split}\label{eq:covprob0}
&\mathcal{P}^{\rm cov}_v
=\mathbb{P}\left \{ \mathcal{C}_{v} \right \}
=\mathbb{P}\left \{ {\rm {SIR}}_v>\tau \right \} .
\end{split}
\end{equation}
As discussed in Sec.~\ref{sec:sys} and \ref{subsec:channel}, the typical user is associated with a single BS in a time-frequency resource block and there are 8 types of BSs in the cellular network. Therefore, the event $\mathcal{C}_{v}$ can be decomposed into 8 sub-events that the typical $v$-user is in coverage when {served} by a BS in $\Psi_b$ with horizontal distance $R_b$. The sub-event is denoted by $\mathcal{B}_{v,b}$.
Therefore, the coverage probability can be rewritten as 
\begin{equation}\label{eq:covprob0_vb}
\begin{split}
\mathbb{P}\left \{ \mathcal{C}_{v} \right \}
&\overset{(a)}{=}\sum_{b\in W_v} \mathbb{E}_{R_b}\left [ \mathbb{P}\left \{ \mathcal{C}_{v},\mathcal{B}_{v,b}|R_b \right \} \right ] 
=\sum_{b\in W_v} \mathbb{E}_{R_b}\left [ \mathbb{P}\left \{ \mathcal{C}_{v}|\mathcal{B}_{v,b} ,R_b\right \}\mathbb{P}\left \{\mathcal{B}_{v,b}|R_b \right \} \right ]\\ 
&=\sum_{b\in W_v}\mathbb{E}_{R_b}\left [ \mathbb{P}\left \{ {\rm SIR}_v^b>\tau |R_b\right \}\mathbb{P}\left \{\mathcal{B}_{v,b}|R_b \right \}\right ],
\end{split}
\end{equation} 
where (a) follows the law of total probability, $R_b$ is given in \eqref{eq:r0}, and ${\rm {SIR}}_v^b$ is given in (\ref{eq:sinrvb}).


\section{Performance Analysis}\label{sec:analy}
This section provide several steps to derive the expressions of the performance metrics defined in Sec.~\ref{sec:sys}. 
First, we derive the distribution of the closest distance between the typical user and each type of BSs. Then, considering that the typical user is associated with a BS with a specific type, we analyze the locations of the nearest interfering BSs for the rest of BS types, followed by the corresponding association probability. We next characterize the interference by its Laplace transform. Finally, we obtain the exact and approximate expressions of coverage probabilities.


\subsection{Distance Distribution}\label{subsec:cdf}
As discussed in Sec.~\ref{subsec:channel}, 
the terrestrial BSs are divided into 8 types, which forms 8 independent and non-homogeneous PPPs: $\Psi_{\rm UML},\Psi_{\rm UMN},\Psi_{\rm USL},\Psi_{\rm USN}, \Psi_{\rm DML}, \Psi_{\rm DMN}, \Psi_{\rm DSL}$, and $\Psi_{\rm DSN}$. The densities of these non-homogeneous PPPs are related to the LoS probability of the channel between a BS and the typical user, e.g. $\lambda_{\rm UML}(r)=\lambda_{\rm U}\mathcal{P}_v^{\rm L}(r)=\delta\lambda_{\rm T}\mathcal{P}_v^{\rm L}(r)$.
With a given mainlobe beamwidth of BSs, the up-tilted BSs that can provide mainlobe gain for the typical aerial (or ground) user is limited in a ring area with a radius range $(z_{a,\rm{U},1},z_{a,\rm{U},2}]$ (or $(z_{g,\rm{U},1},z_{g,\rm{U},2}]$), while the down-tilted BSs that can provide mainlobe gain for the typical aerial (or ground) user is limited in a ring area with a radius range $(z_{a,\rm{D},1},z_{a,\rm{D},2}]$ (or $(z_{g,\rm{D},1},z_{g,\rm{D},2}]$).
The following lemma provides the distribution of the distance between the closest BS in $\Psi_b$ and the typical $v$-user, which is useful to describe event $\mathcal{B}_{v,b}$ that the typical user is associated with different types of BSs.

\begin{lemma}[{Distance Distribution}]\label{lemma1}
\textcolor{black}{
The PDF of the horizontal distance between the typical $v$-user and the closest $b$-{\rm BS} is denoted {by} $f_{v,R_{b}}(r)$, $b=b_1b_2b_3\in W_v$.
For the {up-tilted/down-tilted {\rm BS}s transmitting} signals from mainlobe {with} {LoS/NLoS} links (i.e., $b_1\in\{\rm U,D\}$, $b_2={\rm M}$, and $b_3\in\{\rm L,N\}$), $f_{v,R_{b_1{\rm M}b_3}}(r)$ is given by
\begin{equation}\label{eq:fUML}
f_{v,R_{b_1{\rm M}b_3}}(r)=
\begin{cases}
0\hfill& {\rm if}~r\le z_{v,b_1,1}\\
2\pi\lambda_{b_1}r\mathcal{P}_v^{b_3}(r)\exp\left ( -2\pi\lambda_{b_1}\int_{z_{v,b_1,1}}^{r} z\mathcal{P}_v^{b_3}(z)\mathrm{d}z \right )& {\rm if}~ z_{v,b_1,1}<r\le z_{v,b_1,2}  \\
0\hfill&  {\rm otherwise,}
\end{cases}
\end{equation}
where $z_{v,b_1,j}$ ($j\in\left \{ 1,2 \right \} $) is the maximum or minimum radius of the occurrence area of BSs providing mainlobe gain and is given in (\ref{eq:gainup}) and  (\ref{eq:gaindown}).
For the {up-tilted/down-tilted {\rm BS}s transmitting} signals from sidelobe {with} {LoS/NLoS} links (i.e., $b_1\in\{\rm U,D\}$, $b_2={\rm S}$, and $b_3\in\{\rm L,N\}$), $f_{v,R_{b_1{\rm S}b_3}}(r)$ is given by
\begin{equation}\label{eq:fDSN}
f_{v,R_{b_1{\rm S}b_3}}(r)\!=\!\!
\begin{cases}
2\pi\lambda_{b_1}r\mathcal{P}_v^{b_3}(r)\!\exp\!\!\left (\!\! -\!2\pi\lambda_{b_1}\!\!\int\limits_{0}^{r} \!\!z\mathcal{P}_v^{b_3}(z)\mathrm{d}z \right )\!\! &\!\!{\rm if}~r\!\le\! z_{v,b_1,1}\\
0 \!\!&\!\!{\rm if}~z_{v,b_1,1}\!<\!r\!\le\! z_{v,b_1,2} \\
2\pi\lambda_{b_1}r\mathcal{P}_v^{b_3}(r)\!\exp\!\!\left ( \!\!-\!2\pi\lambda_{b_1} \!\!\left ( \int\limits_{z_{v,b_1,2}}^{r} \!\!\!\!\!z\mathcal{P}_v^{b_3}(z)\mathrm{d}z\!+\!\!\!\!\!\!\int\limits_{0}^{z_{v,b_1,1}} \!\!\!\!\!\!z\mathcal{P}_v^{b_3}(z)\mathrm{d}z \!\!\right ) \!\! \right )\!\! &\!\!{\rm otherwise}.
\end{cases}
\end{equation}
\begin{proof}
See Appendix~\ref{app:lemma1}.
\end{proof}
}
\end{lemma}


\subsection{Nearest Interfering {\rm BS}s}\label{subsec:nearestBS}
Based on the association policy mentioned in Sec.~\ref{subsec:asso}, the typical user is associated with the BS that can provide the strongest average received power instead of the closest BS in the cellular network. Namely, it is impossible for the interfering BSs to provide larger average received power than the serving BS.
From \eqref{eq:r0}, we notice that,
once the horizontal distance between the typical $v$-user and its serving $b$-BS is determined (i.e., $r_0$), the nearest interfering BS in each BS type with distance $R_w$ to the origin is restricted. The minimum value of $R_w$ related to $r_0$ is denoted by $r_{v,w|b}(r_0)$, which is given in the following lemma.

\begin{lemma}[{Nearest Distance of Interfering BSs}]\label{lemma2}
\textcolor{black}{
The typical $v$-user is associated with the {\rm BS} in $\Psi_b$ with horizontal distance $r_0$, implying that the horizontal distance between the interfering {\rm BS}s in $\Psi_w$ ($w\in W_v \setminus \{b\}$) and the typical $v$-user is no less than $r_{v,w|b}(r_0)$, which is given by
\begin{equation}\label{eq:rwb}
r_{v,w|b}(r_0) =
\begin{cases}
\sqrt{(\frac{P^{\rm t}G_{w_2}\eta_{w_3}}{P^{\rm t}G_{b_2}\eta_{b_3}} )^{\frac{2}{\alpha_{w_3}}}\left [ r_0^2+(h_v-h_{\rm T})^2 \right ]^\frac{\alpha_{b_3}}{\alpha_{w_3}}- (h_v-h_{\rm T})^2} 
 & \text{\rm if} ~ r_0^2<X\\
0 & \text{\rm otherwise},
\end{cases}
\end{equation}
where $X=\left [ (\frac{P^{\rm t}G_{w_2}\eta_{w_3}}{P^{\rm t}G_{b_2}\eta_{b_3}} )^{-\frac{2}{\alpha_{w_3}}}(h_v-h_{\rm T})^2 \right ]
^\frac{\alpha_{w_3}}{\alpha_{b_3}} -(h_v-h_{\rm T})^2$.
%
\begin{proof}
The average received power from the serving $b$-BS with distance $r_0$ can be calculated by (\ref{eq:powerrec}) and (\ref{eq:r0}) as $\bar{P}_{v,b}^{\rm r}(r_0)=P^{\rm t}G_{b_2}\zeta_{v,b_3}\left ( r_0 \right )$. Clearly, the interfering {\rm BS}s with type $w$ provide no more average received power than the serving {\rm BS} does. Therefore,  the nearest interfering {\rm BS} in $\Psi_w$ with distance $R_w$ to the typical user satisfies the inequality as follows,
\begin{align}\label{eq:inequality}
P^{\rm t}G_{w_2}\zeta_{v,w_3}\left ( R_w \right ) \le P^{\rm t}G_{b_2}\zeta_{v,b_3}\left ( r_0 \right ),
\end{align}
where $r_{v,w|b}(r_0)$ is the minimum value of $R_w$ in \eqref{eq:inequality}.
However, it might happen that no {\rm BS} in $\Psi_w$ can provide the greater average received power than that by the serving BS with a distance of $r_0$ to the typical user, leading to imaginary $r_{v,w|b}(r_0)$. Naturally, in this case, there is no restriction on the location of the nearest interfering BS in $\Psi_w$, i.e., the minimum value of $R_w$ is $0$. Thus, we complete the proof Lemma~\ref{lemma2}.
\end{proof}
}
\end{lemma}

\subsection{Association Probability}\label{subsec:assoprob}
In Sec.~\ref{subsec:metrics}, the calculation of $\mathcal{P}^{\rm cov}_v$ is transformed into the calculation of the {average joint probability, i.e., $\mathbb{E}_{R_b}\left [ \mathbb{P}\left \{ \mathcal{C}_{v},\mathcal{B}_{v,b}|R_b \right \} \right ]$, that the typical $v$-user is in coverage when served by the nearest BS in $\Psi_b$ with distance $R_b$}. 
In this subsection, we provide the association probability, $\mathcal{A}_{v,b}$, which is the corresponding probability of the event $\mathcal{B}_{v,b}$.

\begin{lemma}[Association Probability]\label{lemma3}
The probability that the typical $v$-user is associated with the nearest {\rm BS} in $\Psi_b$ with distance $r_0$ is given by
\begin{align}\label{eq:assoprob}
\mathcal{A}_{v,b}(r_0)=\xi_{v,b_1b_2}(r_0)\prod_{w\in W\setminus \left \{ b \right \} }\int_{r_{v,w|b}(r_0)}^{\infty } f_{v,R_w}(z)\mathrm{d}z,
\end{align}
where $r_{v,w|b}(r_0)$ is given in (\ref{eq:rwb}), $f_{v,R_w}(z)$ is given in Lemma~\ref{lemma1}, $\xi_{v,b_1b_2}(r)$ is a rectangular function, which is related to the antenna direction ({up-tilt or down-tilt}) i.e., $b_1\in\{\rm U,D\}$, and the antenna gain (mainlobe or sidelobe) of the {\rm BS}s, i.e., $b_2\in\{\rm M,S\}$. The specific expression of $\xi_{v,b_1b_2}(r)$ is defined as
\begin{equation}\label{eq:rect}
\begin{matrix}
 \xi_{v,\rm UM}(r)=&\left\{\begin{matrix}
 1 & \text{\rm if} ~ z_{v,{\rm U},1}<r\le z_{v,{\rm U},1}\\
 0 & \text{\rm otherwise},\hfill
\end{matrix}\right. 
& \xi_{v,\rm DM}(r)=&\left\{\begin{matrix}
 1 & \text{\rm if} ~ z_{v,{\rm D},1}<r\le z_{v,{\rm D},2}\\
 0 & \text{\rm otherwise},\hfill
\end{matrix}\right.\\
 \xi_{v,\rm US}(r)=&1-\xi_{v,\rm UM}(r),\hfill
& \xi_{v,\rm DS}(r)=&1-\xi_{v,\rm DM}(r).\hfill
\end{matrix}
\end{equation}

\begin{proof}
See Appendix~\ref{app:lemma3}.
\end{proof}
\end{lemma}
Note that $z_{\rm{a, D},1}=0$ and $z_{{\rm a,D},2}=0$. 
Therefore, $\xi_{\rm a,DM}(r)$ always equals to $0$, i.e.,  $\mathcal{A}_{{\rm a},\rm DML}(r_0)$ and $\mathcal{A}_{{\rm a},\rm DMN}(r_0)$ are $0$ at any value of $r_0$. Namely, aerial users would never be associated with {\rm DML-BS}s and {\rm DMN-BS}s, which is consistent with our previous results of simplifying the {\rm BS} set $W$ into $W_{\rm a}$ in Sec.~\ref{subsec:channel}. Similarly, for ground users, the association probabilities for {\rm UML-BS}s and {\rm UMN-BS}s are $0$.

\subsection{Interference}\label{subsec:L}

Given the type and the location of the serving BS,
we characterize the interference at the typical $v$-user, i.e., $I_{v|r_0}$ in \eqref{eq:interference}, by its Laplace transform~\cite{andrews2016primer}.

\begin{lemma}[Laplace Transform of Interference]\label{lemma4}
The Laplace transform of the interference (conditioned on the type of the serving {\rm BS} being $b$ and the horizontal distance between the serving BS and the user being $r_0$), denoted  by $\mathcal{L}_{I_{v|r_0}}(s)$, is given by
{
\begin{equation}\label{eq:L}
\begin{split}
\mathcal{L}_{I_{v|r_0}}(s)
&=\exp\left (-\sum_{w\in W_v}2\pi \lambda_{w_1}\int_{r_{v,w|b}(r_0)}^{\infty }[1-\kappa_{w}(z,s)]z\mathcal{P}_v^{ w_3}(z)\xi_{v,w_1w_2}(z)\mathrm{d}z\right ),
\end{split}
\end{equation}
}
where $\kappa_{w}(z,s)=(\frac{m_{w_3}}{m_{w_3}+sP^{\rm t}G_{w_2}\zeta_{v,w_3}(z)} )^{m_{w_3}}$ and if $w=b$, $r_{v,w|b}(r_0)=r_0$.
\begin{proof}
See Appendix~\ref{app:lemma4}.
\end{proof}
\end{lemma}

\subsection{Exact Coverage Probability}\label{subsec:covprob1}
As discussed in Sec.~\ref{subsec:metrics}, the system coverage probability is equivalent to the sum of coverage probabilities conditioned that the typical user is respectively associated with 8 types of BSs.
Based on the association probability and {the distribution of the distance} between the typical user and its serving BS, in this subsection, we derive the expression of $\mathcal{P}^{\rm cov}_v$.
\begin{lemma}[Exact Coverage Probability]\label{lemma5}
The coverage probability of the typical $v$-user in a cellular network containing both {up-tilted {\rm BS}s and down-tilted {\rm BS}s} with the average received power-based association policy, i.e., $\mathcal{P}^{\rm cov}_v$, is given by
\begin{equation}\label{eq:covprob1}
\begin{split}
\mathcal{P}^{\rm cov}_v
&=\sum_{b\in W_v}\int_{0}^{\infty} \mathcal{P}^{\rm cov}_{v,b}(r_0)\mathcal{A}_{v,b}(r_0)f_{v,R_b}(r_0)\mathrm{d}r_0 ,
\end{split}
\end{equation}
where $\mathcal{A}_{v,b}(r_0)$ is given in (\ref{eq:assoprob}), $f_{v,R_b}(r_0)$ is given in Lemma~\ref{lemma1}, and $\mathcal{P}^{\rm cov}_{v,b}(r_0)$ is the conditional coverage probability given that the typical $v$-user is associated with the $b$-type {\rm BS} at horizontal distance $r_0$. From \eqref{eq:covprob0_vb}, $\mathcal{P}^{\rm cov}_{v,b}(r_0)=\mathbb{P}\left \{ {\rm SIR}_v^b>\tau |r_0\right \}$, which can be further expressed as
\begin{equation}
\begin{split}\label{eq:covprob_vb}
\mathcal{P}^{\rm cov}_{v,b}(r_0)=\sum_{k=0}^{m_{b_3}-1} \frac{(-s)^k}{k!}\frac{\mathrm{d^k} }{\mathrm{d^k} s} {\mathcal{L}_{I_{v|r_0}}(s)} ,
\end{split}
\end{equation}
where $s=\frac{m_{b_3}\tau}{P^{\rm t}G_{b_2}\zeta_{v,b_3}(r_0)}$ and ${\mathcal{L}_{I_{v|r_0}}(s)}$ is given in (\ref{eq:L}).

\begin{proof}
See Appendix~\ref{app:lemma5}.
\end{proof}
\end{lemma}

\subsection{Approximate Coverage Probability}\label{subsec:covprob2}
When $m_{b_3}$ is larger than $1$, the calculation of $\mathcal{P}^{\rm cov}_{v,b}(r_0)$ in (\ref{eq:covprob_vb}) would be quite complex due to the high order of derivations of the Laplace transform ${\mathcal{L}_{I_{v|r_0}}(s)}$. Therefore, we provide an approximate expression for $\mathcal{P}^{\rm cov}_{v,b}(r_0)$ to simplify the computation in the following. 
\begin{lemma}[Approximate Coverage Probability]\label{lemma6}
Using the upper bound of {\rm CDF} of the Gamma distribution, the complex expression of $\mathcal{P}^{\rm cov}_v$ can be approximated as~\cite{approximate}
\begin{equation}
\begin{split}\label{eq:approxi_covprob0}
\tilde{\mathcal{P}} ^{\rm cov}_v
=\sum_{b\in W_v}\int_{0}^{\infty} \sum_{k=1}^{m_{b_3}}\binom{m_{b_3}}{k}\left ( -1 \right )^k {\mathcal{L}_{I_{v|r_0}}}\left ( k \beta_{b_3} s \right )
\mathcal{A}_{v,b}(r_0)f_{v,R_b}(r_0)\mathrm{d}r_0,
\end{split}
\end{equation}
where $\beta_{b_3}=\left ( m_{b_3}! \right ) ^{\frac{-1}{m_{b_3}} }$, $s$ is given in (\ref{eq:covprob_vb}) and  ${\mathcal{L}_{I_{v|r_0}}}\left ( k \beta_{b_3} s \right )$ is obtained by {using} (\ref{eq:L}).
\begin{proof}
See Appendix~\ref{app:lemma6}.
\end{proof}
\end{lemma}


\section{Results and Discussion}\label{sec:simu}
In this section, we provide the numerical results and extensive Monte Carlo simulation results. To verify the effectiveness of the proposed method,
we compare the coverage probabilities of aerial users and ground users in a network consisting of up-tilted BSs and down-tilted BSs with that in the current network without {up-tilted} BSs. Furthermore, we explore the effects of the system parameters, which allows the network designer to improve the connectivity of both aerial users and ground users. The parameters used in the simulations and their default values are given in {Table}.~\ref{tab:simulation}, unless otherwise specified.

\begin{table}[t!]\caption{Table of System Numerical Parameters.}
\centering
\begin{center}
{
\renewcommand{\arraystretch}{1.2}\small
    \begin{tabular}{ | {c} | {c} || {c} | {c} | }
    \hline
        \hline
    \textbf{System Parameters} & \textbf{Default Values} &  \textbf{System Parameters} & \textbf{Default Values} \\ \hline
    $(h_{\rm T},h_{\rm a},h_{\rm g})$ & $(50, 100, 0) ~\rm m$ & $\lambda_{\rm T}$ & $2\times 10^{-6} ~\rm{BSs/m^2}$\\ \hline
    $P^{\rm t}$ & $-6~\rm dB$ & $(G_{\rm M},G_{\rm S})$ & $(10, 0.5) ~\rm dB$\\ \hline
    $(\theta_{\rm U},\theta_{\rm D})$ & $(12, 12) ^{\circ}$ & $(\varphi_{\rm U},\varphi_{\rm D})$ & $(20, 20) ^{\circ}$  \\ \hline
          $(m_{\rm L},m_{\rm N})$ & $(1, 3)$ &$(\alpha_{\rm L},\alpha_{\rm N})$ & $(2.09, 3.75)$\\ \hline
          $(\eta_{\rm L},\eta_{\rm N})$ & $(-41.1, -32.9) ~\rm dB$ & $\tau$ & $-5 ~\rm dB$ \\ \hline
    \end{tabular}}
\end{center}
\label{tab:simulation}
\end{table}

\subsection{Impact of the Fraction of {Up-tilted} BSs and Communication Environment} \label{subsec:delta}

\begin{figure}%
 \centering
 \subfloat[]{\includegraphics[width=0.5\columnwidth]{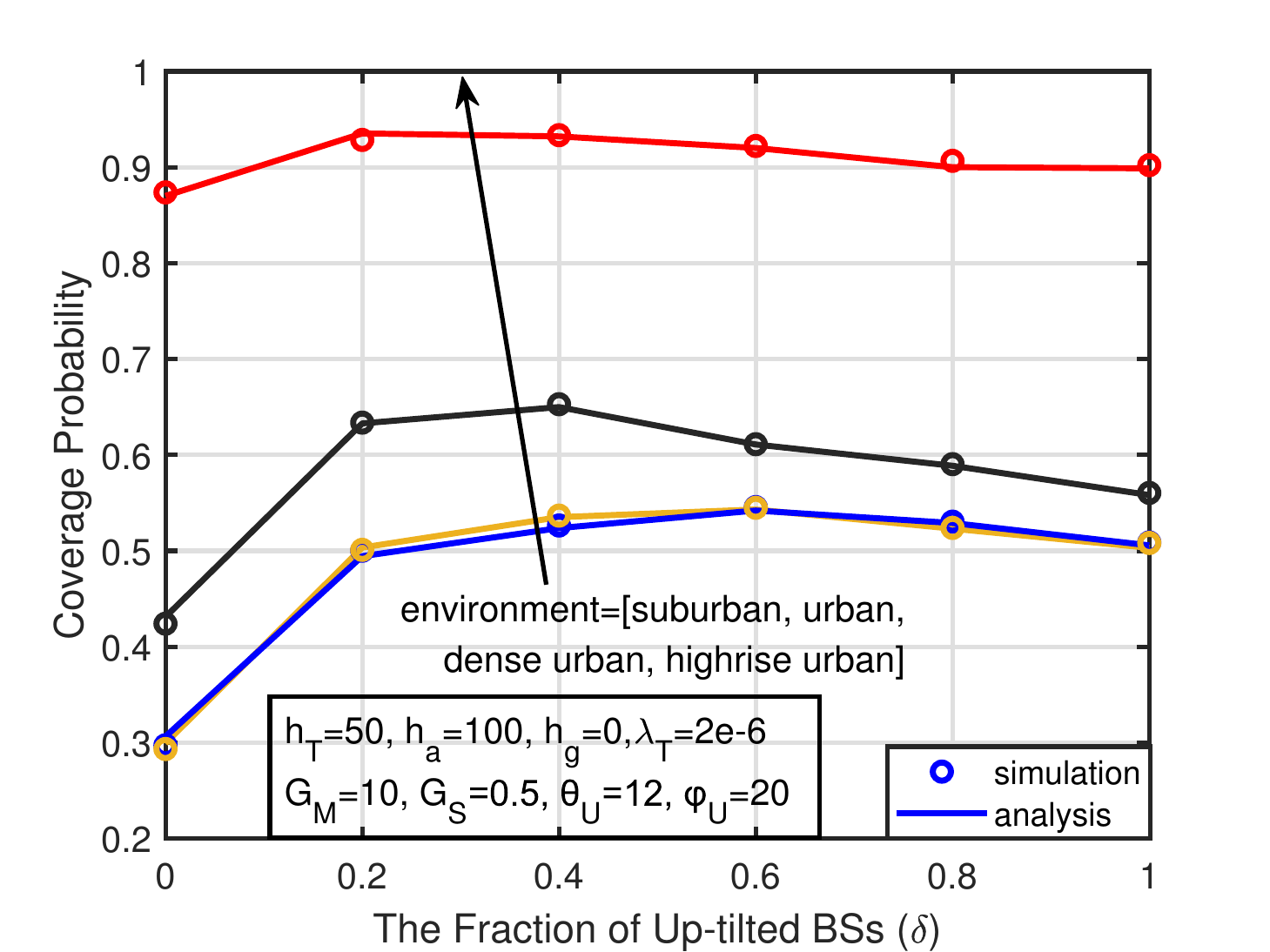}\label{fig:delta_envir1}}
 \subfloat[]{\includegraphics[width=0.5\columnwidth]{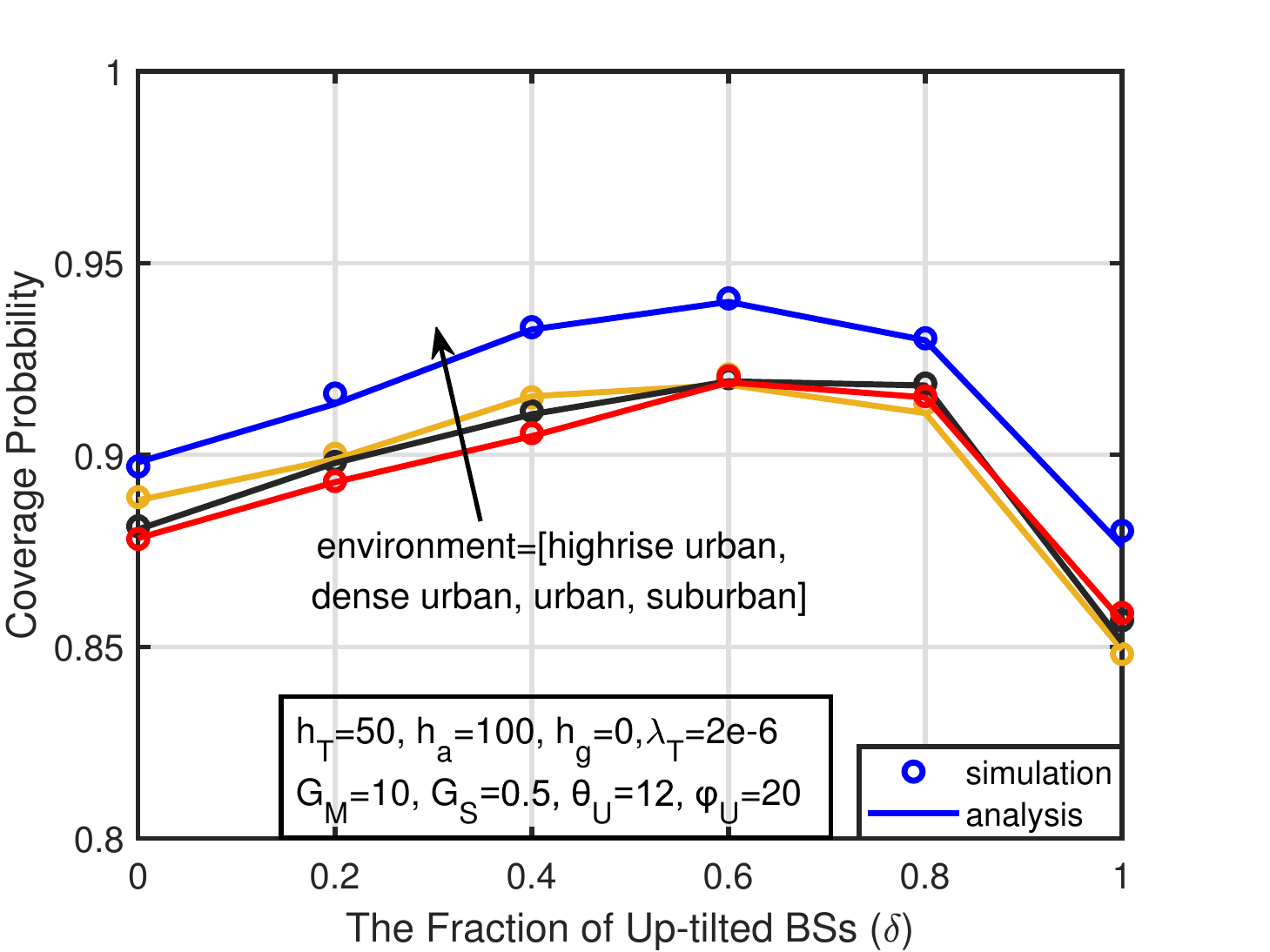}\label{fig:delta_envir2}}
 \caption{Comparing with $\tau=-5~{\rm dB}$, the value of $\mathcal{P}^{\rm cov}$ vs the fraction of up-tilted BSs ($\delta$) in different environments for (a) aerial users, (b) ground users. }%
 \label{fig:delta_envir} 
\end{figure}

In Fig.~\ref{fig:delta_envir}, we plot the coverage probability vs the fraction of up-tilted BSs ($\delta$) in four selected environments~\cite{environment}: suburban $(0.1, 750, 8)$, urban $(0.3, 500, 15)$, dense urban $(0.5, 300, 20)$, and highrise urban $(0.5, 300, 50)$ for aerial users and ground users, respectively.
The simulation results closely match the numerical results of \eqref{eq:approxi_covprob0}, which verifies the accuracy of our analysis.  

{\em Feasibility of the proposed method.}
It is worth mentioning that the statistics observed at $\delta=0$ corresponds to the performance of the current cellular network composed entirely of {down-tilted} BSs.
It can be seen from Fig.~\ref{fig:delta_envir1} that the value of the aerial coverage probability when $0<\delta\le 1$ (i.e., in the proposed network) is greater than that when $\delta=0$ (i.e., in the current network), indicating that steering part of the BSs antennas into users in the sky effectively improve the QoS of aerial users.
Interestingly, as shown in Fig.~\ref{fig:delta_envir2}, the occurrence of {up-tilted} BSs also improves the {QoS} of ground users, as long as the fraction of {up-tilted} BSs is moderate, approximately between $0.1$ and $0.9$. 
These observations clarify the feasibility of the proposed cellular network, which not only ensures the connectivity of aerial users but also increases the communication quality of the ground users.

{\em The fraction of up-tilted BSs.}
In Fig.~\ref{fig:delta_envir1}, we see that the aerial coverage probability in the dense urban environment achieves its maximum value when we increase $\delta$ to $0.4$; while a further increase in $\delta$ lowers the coverage performance. 
It is clear that, as $\delta$ increases, aerial users have a higher chance to be associated with an up-tilted BS and receive power through its mainlobe, thereby improving the coverage probability.
However, despite the high received power from the serving BS, the rest of {up-tilted} BSs interfere with the aerial users from the mainlobes/sidelobes. This is why a larger number of up-tilted BSs (e.g., $\delta>0.4$) results in a decrease in the SIR.
In Fig.~\ref{fig:delta_envir2}, we also see that the coverage probability of ground users in the dense urban environment first increases with $\delta$ until $0.8$, but as $\delta$ further increases, the performance begins to decrease.
In fact, for ground users, a suitable ratio of {up-tilted} BS not only converts part of the mainlobe interference into sidelobe interference but also ensures their association with down-tilted BSs that can provide mainlobe gain.
It can be concluded from Fig.~\ref{fig:delta_envir1} and \ref{fig:delta_envir2} that a network consisting entirely of one type of BS, either up-tilted BSs ($\delta=1$) or down-tilted BSs ($\delta=0$), does not satisfy the high QoS of both the aerial users and the ground users. 
In the rest of the simulations, we focus on $\delta=0.4$, at which the performance of the cellular network is enhanced for both aerial users and ground users, and then we discuss the impact of other system parameters.

{\em Communication environment.}
It is also visible from Fig.~\ref{fig:delta_envir1} that, in the suburban, urban, and dense urban areas, the coverage probabilities of aerial users are more sensitive to the value of $\delta$; while highrise urban aerial users have a high coverage probability with slight fluctuations as $\delta$ varies. 
In the highrise urban environment, the high and dense blockages increase the path loss during the signal transmission, greatly reducing the power from interfering BSs. This is why increasing the number of up-tilted BSs has little impact on the highrise urban aerial users. 
In the other three environments, T2A links are generally LoS due to the high altitude of aerial users. Consequently, the interfering BSs interfere with the aerial user via LoS links, especially when the number of up-tilted BSs increases, thereby degrading the communication quality.


\subsection{Impact of {Up-tilt} Angle and {Down-tilt} Angle} \label{subsec:theta}

\begin{figure}%
 \centering
 \subfloat[]{\includegraphics[width=0.5\columnwidth]{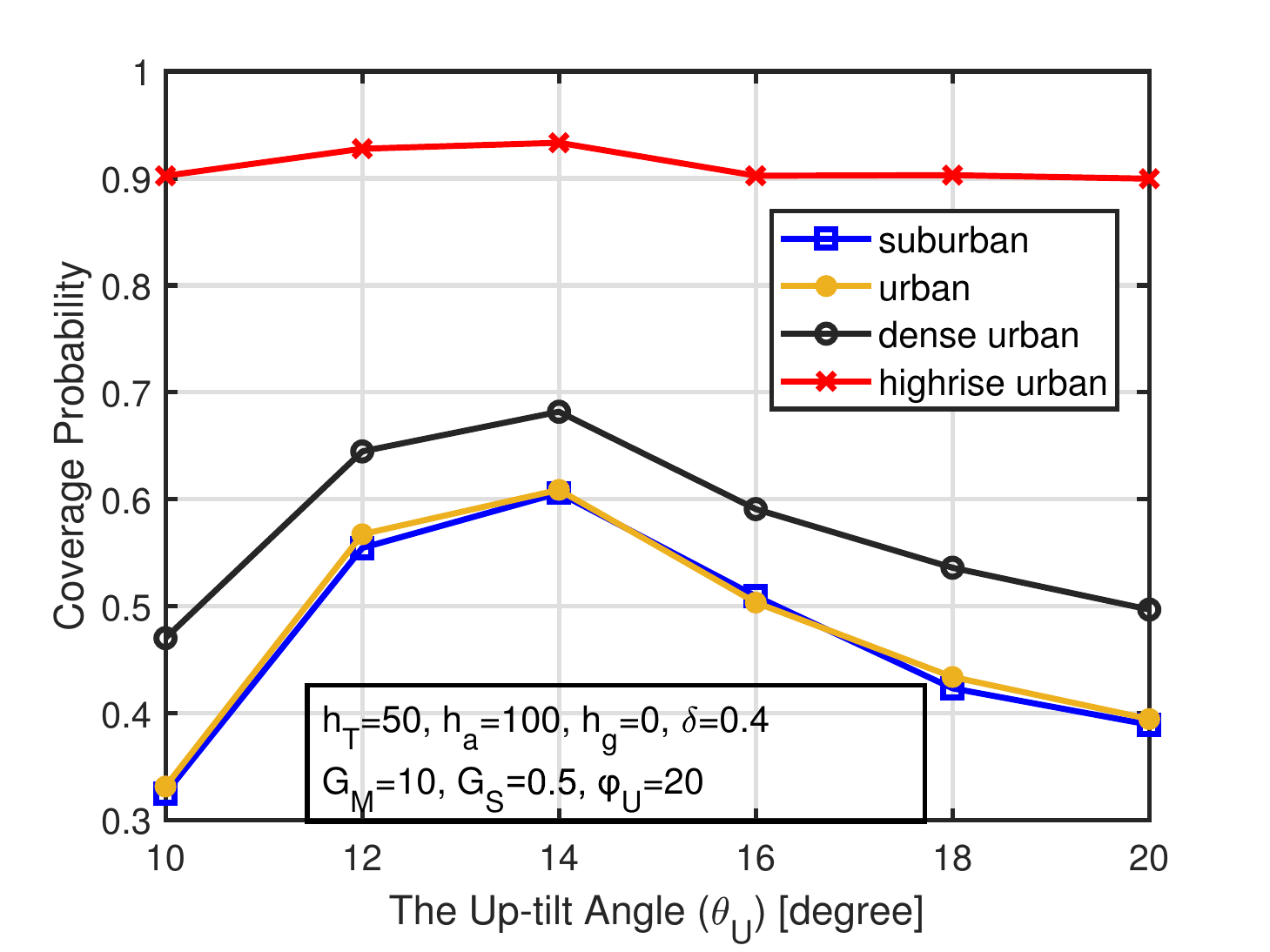}\label{fig:theta_envir1}}
 \subfloat[]{\includegraphics[width=0.5\columnwidth]{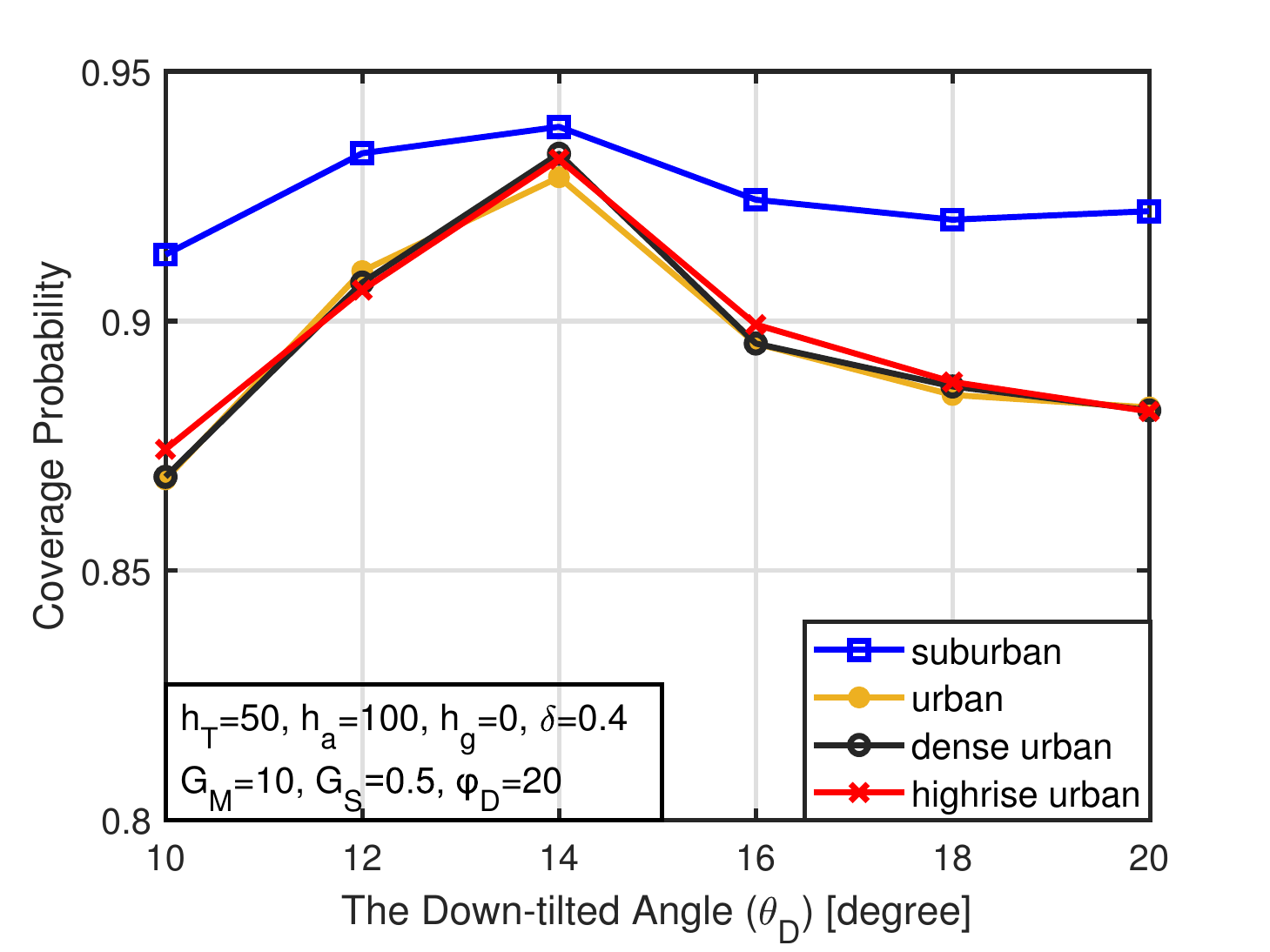}\label{fig:theta_envir2}}
 \caption{Comparing with $\tau=-5~{\rm dB}$, the value of $\mathcal{P}^{\rm cov}$ for different {values} of $\theta_{\rm U}$ or $\theta_{\rm D}$ in different environments for (a) aerial users, (b) ground users. }%
 \label{fig:theta_envir} 
\end{figure}

Fig.~\ref{fig:theta_envir} reveals the impact of the {up-tilt} angle and the {down-tilt} angle on the coverage probabilities of aerial users and ground users, respectively.
As discussed in (\ref{eq:GS}), the down-tilted BSs only serve/interfere with aerial users through sidelobes, implying that the down-tilt angle has no effect on the performance of aerial users. Likewise, the different values of the up-tilt angle do not affect the QoS of the aerial users.
Therefore, we only draw the curve of the coverage probability for aerial users when changing the angle and beamwidth of {up-tilted} antennas in Fig.~\ref{fig:theta_envir1} and do the opposite in Fig.~\ref{fig:theta_envir2}.

{\em Up-tilt Angle.}
From (\ref{eq:gainup}), 
the beamwidth of up-tilted BSs ($\varphi_{\rm U}$) and the up-tilt angle ($\theta_{\rm U}$) define {the mainlobe coverage area} in a horizontal plane, where the inner ring radius or the outer ring radius of the area is $z_{v,{\rm U},j}$ ($j\in\left \{ 1,2 \right \} $). BSs falling in this mainlobe coverage area can provide service through mainlobe for the typical aerial-user. 
We set the beamwidth of {up-tilted} BSs at a fixed value as $\varphi_{\rm U}=20^{\circ}$. In this case, the inner/outer ring radius, i.e., $z_{v,{\rm U},j}$, is determined by the {up-tilt} angle. 
We notice that when $\varphi_{\rm U}=20^{\circ}$ and $\theta_{\rm U}=10^{\circ}$, 
$z_{v,{\rm U},2}=\infty$. Namely, most interference comes from the mainlobe of UML-BSs and UMN-BSs, thereby limiting the {SIR}. This explains the worst coverage performance at the first point in Fig.~\ref{fig:theta_envir1}.
When $\theta_{\rm U}$ increases from $10^{\circ}$ to $14^{\circ}$, the mainlobe coverage area is gradually shrinking, i.e., the number of UML-BSs and UMN-BSs decreases. Correspondingly, the interference reduces
to reduce interference. The above analysis is consistent with the initial increasing tendency of the aerial coverage probability in Fig.~\ref{fig:theta_envir1}.
However, Fig.~\ref{fig:theta_envir1} also shows that further up-tilting the BS antennas leads to a decrease in the aerial coverage probability.
As discussed above,
the larger value of $\theta_{\rm U}$, the smaller the mainlobe coverage area, the smaller the number of UML-BSs and UMN-BSs, and even there are no BSs in this {mainlobe coverage area}.
Therefore, the typical aerial-user has to connect with a USL-BS or a DSL-BS that provides sidelobe gain. Obviously, the received power from a USL-BS or a DSL-BS is less than that from a UML-BS under the same propagation distance. 

{\em Down-tilt Angle.}
We see the same trend in Fig.~\ref{fig:theta_envir2} compared to Fig.~\ref{fig:theta_envir1}. The conclusion and explanation for the impact of the up-tilt angle on the aerial coverage probability apply to the impact of the down-tilt angle on the ground coverage probability. Therefore, we omit it here.


\subsection{Impact of Beamwidth and Height of \rm BSs} \label{subsec:phi}

\begin{figure}%
 \centering
 \subfloat[]{\includegraphics[width=0.5\columnwidth]{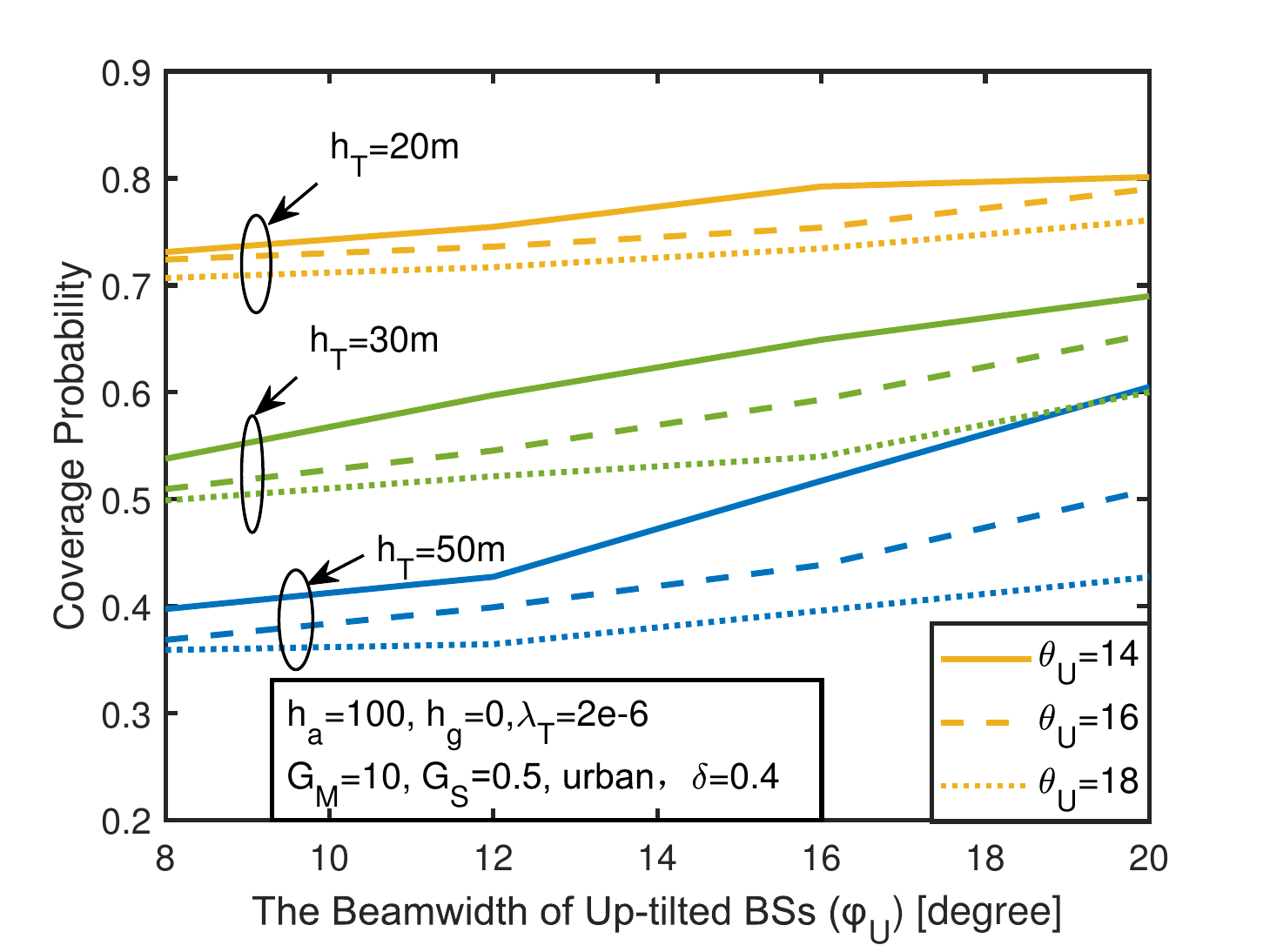}\label{fig:phi1}}
 \subfloat[]{\includegraphics[width=0.5\columnwidth]{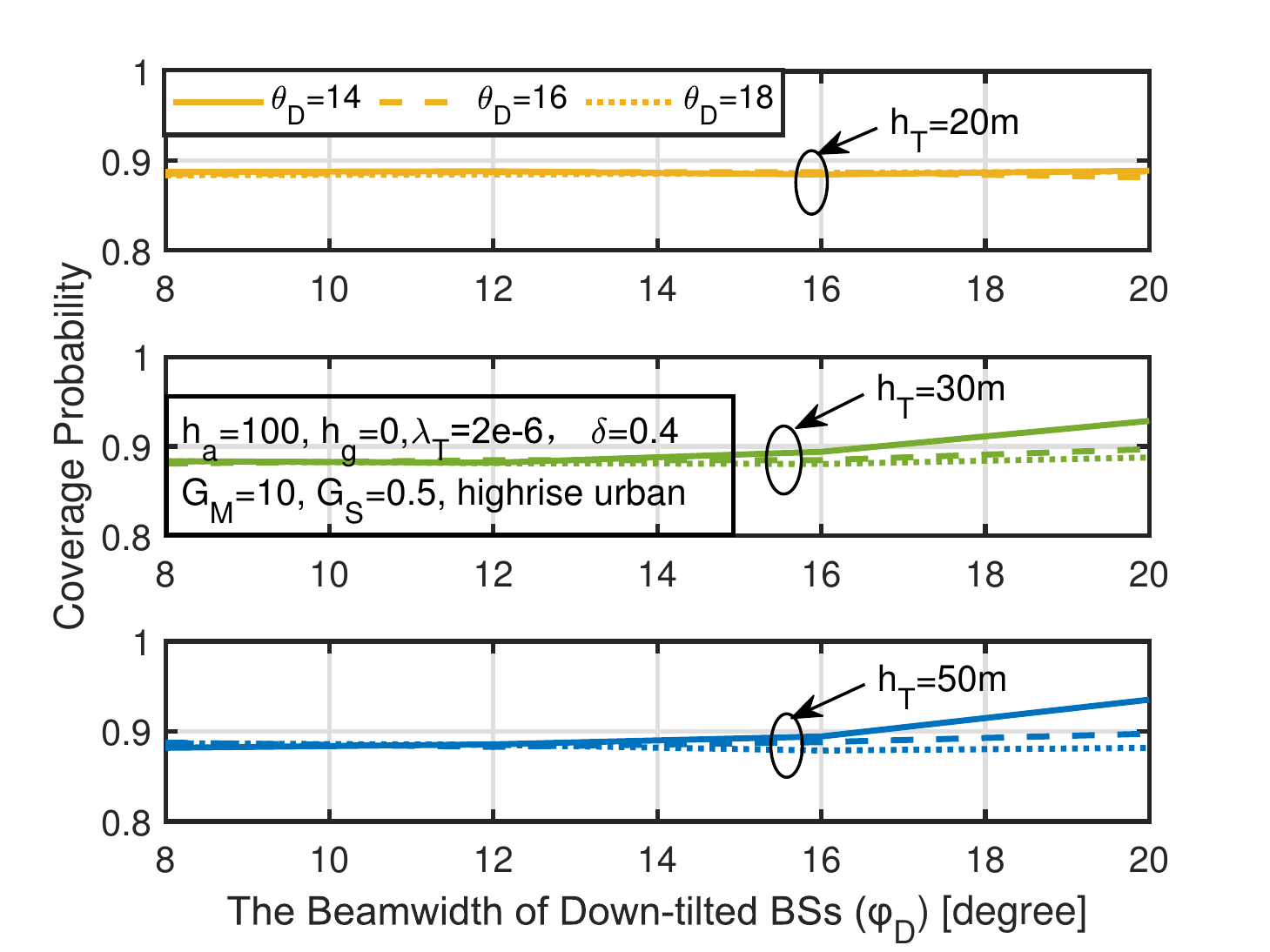}\label{fig:phi2}}
 \caption{Comparing with $\tau=-5~{\rm dB}$, the value of $\mathcal{P}^{\rm cov}$ for different {values} of $\varphi_{\rm U}$ for (a) aerial users under an urban environment, (b) ground users under a highrise environment. }%
 \label{fig:phi} 
\end{figure}

In Fig.~\ref{fig:phi}, we present the coverage probability vs the beamwidth of {up-tilted}/down-tilted BSs for different {up-tilt}/down-tilted angles and different BS heights under an urban environment. 
As discussed in Sec.~\ref{subsec:theta}, varying the value of the up-tilted (or down-tilt) beamwidth only affect the QoS of aerial (or ground) users. 
Therefore, we separately investigate the impact of the up-tilted beamwidth on aerial users in Fig.~\ref{fig:phi1} and the impact of the down-tilt beamwidth on ground users in Fig.~\ref{fig:phi2}.

{\em Beamwidth of up-tilted BSs.}
We see from Fig.~\ref{fig:phi1} that, with the increasing beamwidth of the {up-tilted} BSs, the urban aerial users tend to be better served.
As mentioned in Sec.~\ref{subsec:theta}, the range of the mainlobe coverage area, i.e., $(z_{v,{\rm U},1},z_{v,{\rm U},1}]$, depends on both 
the beamwidth of up-tilted BSs ($\varphi_{\rm U}$) and the up-tilt angle ($\theta_{\rm U}$).
Therefore, the larger $\theta_{\rm U}$, the larger the {mainlobe coverage area}. Namely, 
the chance of the typical aerial user being associated with a UML-BS becomes greater.
Similar to the results in Fig.~\ref{fig:theta_envir}, Fig.~\ref{fig:phi1} also shows that a large value of the {up-tilt} angle (e.g., $\theta_{\rm U}=18^{\circ}$) destroys the {QoS} of aerial users. 
The results shown in Fig.~\ref{fig:phi1} and Fig.~\ref{fig:phi1} provide guidelines for designing parameters of {up-tilted} BSs, i.e., a combination of
a large value of $\varphi_{\rm U}$ and a moderate value of $\theta_{\rm U}$.

{\em Height of up-tilted BSs.}
Fig.~\ref{fig:phi1} intuitively illustrates the influence of BSs height on the aerial coverage probability, i.e. a rise in BSs height makes the coverage performance of the aerial users worse. 
From (\ref{eq:gainup}), (\ref{eq:gaindown}), and (\ref{eq:PLoS}), we see that the height of BSs has a potential influence on the {mainlobe coverage area} and LoS transmission.
In fact, when the BSs are at a higher altitude, the {mainlobe coverage area} in the ground is extended and the probability of LoS transmission is improved, which leads to severe interference and thus reduces the {SIR}. 

{\em Beamwidth and height of down-tilted BSs.}
A similar impact of the beamwidth and height of {down-tilted} BSs is shown in Fig.~\ref{fig:phi2}.


\subsection{Impact of Aerial Users Height } \label{subsec:ha}

\begin{figure}
\centering
\includegraphics[width=0.5\columnwidth]{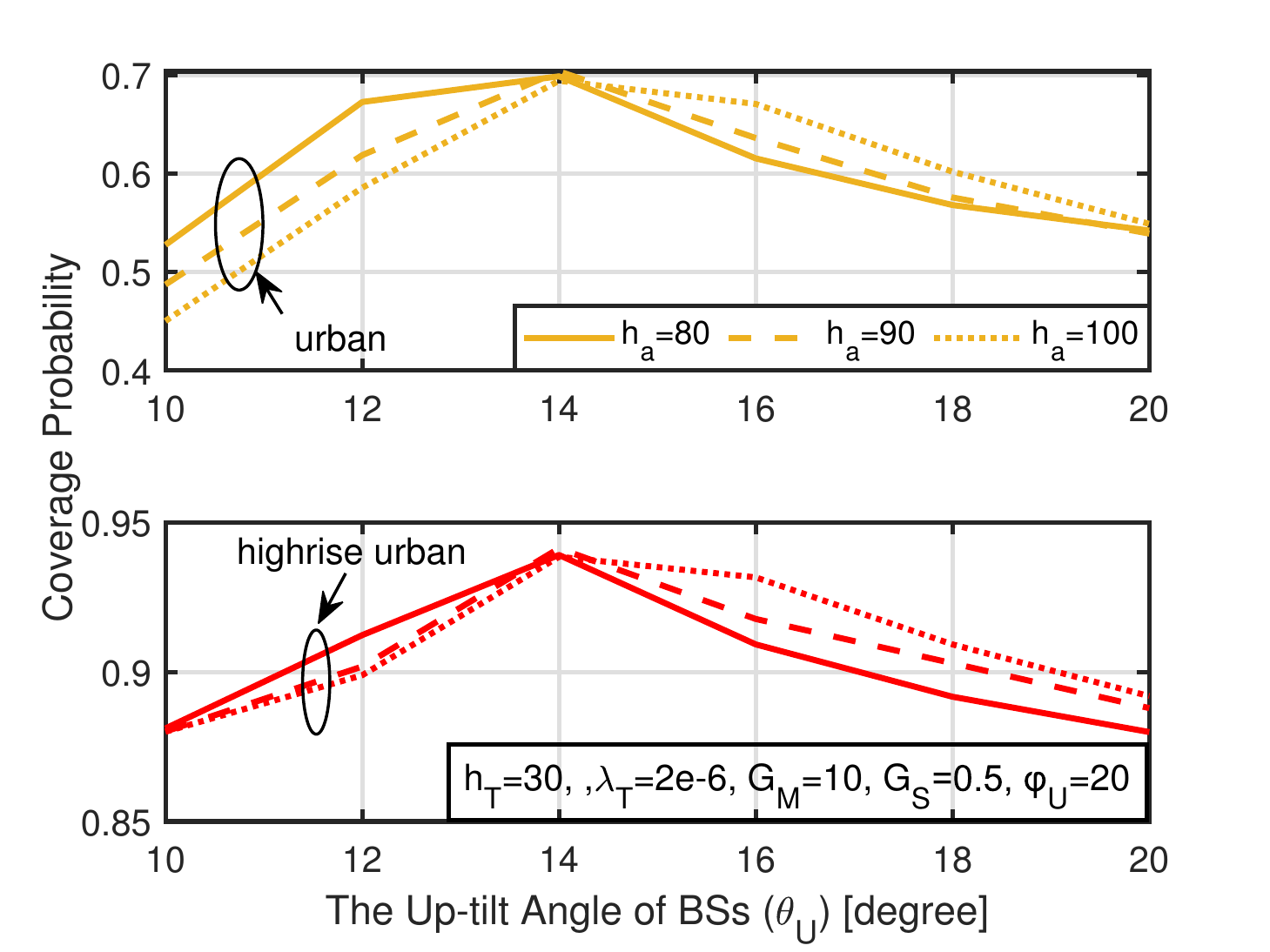}
\caption{ Comparing with $\tau=-5~{\rm dB}$, the value of $\mathcal{P}^{\rm cov}$ for different {values} of $\theta_{\rm U}$ given the height of aerial users and the environment. }
\label{fig:ha_theta_U}
\end{figure}

Fig.~\ref{fig:ha_theta_U} provides some suggestions on how to adjust the {up-tilt} angle given the height of the urban aerial user to ensure connectivity.
For {up-tilted} BSs, in addition to the height of BSs, the size of the {mainlobe coverage area} and the probability of LoS transmission are related to the height of aerial users as well.
In general, the coverage probability decreases as the height of aerial users increases due to the higher probability of LoS transmission and the corresponding stronger interference power. This conclusion is consistent with the simulations results in Fig.~\ref{fig:ha_theta_U} when $\theta_{\rm U}<14^{\circ}$ and $\varphi_{\rm U}=20^{\circ}$. 
However, 
when $\theta_{\rm U}>14^{\circ}$, there is a completely opposite trend of the coverage probability varying with the altitude of aerial users. The reason for this opposite trend is that the propagation distance becomes dominant in the SIR. Therefore, the higher the aerial users are, the better trade-off between the received power and the interference can achieve.


\section{Conclusions}\label{sec:conclusion}
{
In this paper, we proposed an effective method to adjust the current cellular network for aerial users without reducing the QoS of the ground users, i.e., converting a subset of down-tilted BSs into up-tilted BSs.
We also provided a stochastic geometry-based framework for the proposed cellular network.
Specifically, we derived the analytical expressions of the coverage probabilities for aerial users and ground users, respectively.
We observed the improved coverage probabilities for both aerial and ground users in the adjusted cellular network as compared with the current cellular network, which strongly supports the feasibility of the proposed method.
With this framework, we further investigated the effect of system parameters on the coverage performance, including the fraction of up-tilted BSs, the up-tilted angle, the beamwidth of BS antennas, and the height of aerial users/BSs. For instance, we found the optimal value of the fraction of up-tilted BSs for enhancing the coverage performance for both aerial and ground users. 
Therefore, our work allows the network designer to design the future cellular network for satisfying the QoS of aerial users and ground users, before actual deployment.
}

As for future work, some issues that have not been discussed in this paper are still worth investigating, e.g. a more complicated antenna model, the mobility of aerial users, and the characteristics of T2A links.
{\em (i) A more complicated antenna model.}
\textcolor{black}{The scenario where the BS adopts 3D beamforming and the users are equipped with directional antennas to suppress the interference power can be considered when analyzing \cite{lyu2019network,vilor2020optimal}. Interestingly, even under the assumption that the antenna radiation pattern is omnidirectional in the horizontal plane and directional in the vertical plane as described in this paper, the QoS of aerial users and ground users can be enhanced. It is reasonable to deduce that when the BSs and the users adopt a 3D antenna radiation pattern with a more finite mainlobe coverage area, aerial users and ground users can have better coverage performance.}
{\em (ii) The mobility of aerial users.}
\textcolor{black}{In this paper, the BSs with fixed up-tilted antennas are randomly distributed and follow a PPP. However, considering that the aerial users are mostly movable vehicles in the future, instead of fixing the antenna direction of BSs as up-tilt or down-tilt, a new dimension of improving the QoS of aerial users and ground users is worth trying, i.e. deciding which BSs to be up-tilted based on a particular traffic state of the network. The adjustable direction of BS antennas ensures the 
efficient utilization of BSs, thereby enhancing the QoS of both aerial users and ground users.}
{\em (iii) The characteristics of T2A links.} 
Future aerial transportation is expected to change the high LoS probability of the T2A links because the users are densely distributed in different altitudes and the users located between the height of the desired user and the height of the serving BS may block the communication channel. Therefore, the LoS probability of T2A links should be remeasured.

\appendices
\section{Proof of Lemma~\ref{lemma1}}\label{app:lemma1}
In Sec.~\ref{subsec:channel}, we divided a HPPP $\Psi_{\rm T}$ into 8 independent and non-homogeneous PPPs, i.e., $\Psi_{\rm UML}$, $\Psi_{\rm UMN}$, $\Psi_{\rm USL}$, $ \Psi_{\rm USN}$, $\Psi_{\rm DML}$, $\Psi_{\rm DMN}$, $ \Psi_{\rm DSL}$, and $\Psi_{\rm DSN}$.
We start by analyze $\Psi_{\rm UML}$. From \eqref{eq:gainup},
the {up-tilted} BSs that can provide mainlobe gain are restricted to the ring area with a radius ranging from $z_{v,{\rm U},1}$ to $z_{v,{\rm U},2}$. Thus, the Lebesgue measure of the UM-BSs area, denoted by $\rho_{v,\rm UM}$,
centered at the origin with a radius of $r$ can be expressed as~\cite{stochasticgeometry}
\begin{equation}\label{eq:pUML}
\rho_{v,\rm UM}(r)=
\begin{cases}
0\hfill& {\rm if}~r\le z_{v,{\rm U},1}\\
\pi(r^2-{z_{v,{\rm U},1}}^2)& {\rm if}~z_{v,{\rm U},1}<r\le z_{v,{\rm U},2}  \\
\pi({z_{v,{\rm U},2}}^2-{z_{v,{\rm U},1}}^2)& {\rm otherwise.}
\end{cases}
\end{equation}
$R_{\rm UM}$ is the nearest horizontal distance of UM-BSs to the origin.
Using the null probability of the PPP in~\cite{nullprob}, the cumulative distribution function (CDF) of $R_{\rm UM}$ is given by
%
\begin{equation}\label{eq:FUM}
\begin{split}
F_{v,R_{\rm UM}}(r)&=\mathbb{P}\left \{ R_{\rm UM}\le r \right \} 
=1-\mathbb{P}\left \{ R_{\rm UM}>r \right \}
=1-\exp\left [ \lambda_{\rm U}\rho_{v,\rm UM}(r) \right ] \\  &=
\begin{cases}
0\hfill &{\rm if}~r\le z_{v,{\rm U},1}\\
1-\exp\left ( -\lambda_{\rm U}\pi(r^2-{z_{v,{\rm U},1}}^2) \right ) &{\rm if}~ z_{v,{\rm U},1}<r\le z_{v,{\rm U},2}  \\
1-\exp\left ( -\lambda_{\rm U}\pi({z_{v,{\rm U},2}}^2-{z_{v,{\rm U},1}}^2) \right ) &{\rm otherwise.}\\
\end{cases}
\end{split}
\end{equation}
Taking the probability of LoS transmission in (\ref{eq:PLoS}) into account, the density of UML-BSs is $\lambda_{\rm ULM}(r)=\lambda_{\rm U}\mathcal{P}_v^{\rm L}(r)$ and the CDF of UML-BSs is given by
\begin{equation}\label{eq:FUML}
\begin{split}
F_{v,R_{\rm UML}}(r)=
\begin{cases}
0 \hfill &{\rm if}~r\le z_{v,{\rm U},1}\\
1-\exp\left ( -2\pi\lambda_{\rm U} \int_{z_{v,{\rm U},1}}^{r} z\mathcal{P}_v^{\rm L}(z)\mathrm{d}z \right ) \hfill &{\rm if}~ z_{v,{\rm U},1}<r\le z_{v,{\rm U},2}  \\
1-\exp\left ( -2\pi\lambda_{\rm U} \int_{z_{v,{\rm U},1}}^{z_{v,{\rm U},2}} z\mathcal{P}_v^{\rm L}(z)\mathrm{d}z \right ) &{\rm otherwise.}
\end{cases}
\end{split}
\end{equation}
Considering the relationship between the PDF and the CDF, i.e., $f_{v,R_{\rm UML}}(r)=\frac{\mathrm{d} }{\mathrm{d} r} F_{v,R_{\rm UML}}(r)$, we obtain the expression of $f_{v,R_{\rm UML}}(r)$. The density of $\Psi_{\rm UMN}$ is $\lambda_{\rm UNL}(r)=\lambda_{\rm U}\mathcal{P}_v^{\rm N}(r)$. $f_{v,R_{\rm UMN}}(r)$ can be derived by following the above methods. Likewise, $f_{v,R_{b}}(r)$ can be derived similar to $f_{v,R_{\rm UML}}(r)$ and $f_{v,R_{\rm UMN}}(r)$ and we omit it here.

\section{Proof of Lemma~\ref{lemma3}}\label{app:lemma3}
${A}_{v,b}(r_0)$ is the probability that the typical $v$-user is associated with the nearest BS in $\Psi_b$ with horizontal distance $r_0$. 
From \eqref{eq:r0},
the average received power from the serving BS in $\Psi_b$ is larger than that from the closest BS with distance $R_w$ in $\Psi_w$ ($ w\in W_v\setminus \left \{ b \right \} $), i.e., $\bar{P}_{v,b}^{\rm r}(r_0)>\bar{P}_{v,w}^{\rm r}(R_w)$. Hence, ${A}_{v,b}(r_0)$ can be expressed as
\begin{equation}\label{eq:assoprob1}
\begin{split}
\mathcal{A}_{v,b}(r_0)&=\xi_{v,b_1b_2}(r_0)\mathbb{P}\left \{ \mathcal{B}_{v,b}(r_0) \right \}=\xi_{v,b_1b_2}(r_0)\prod_{w\in W_v\setminus \left \{ b \right \} } \mathbb{P}\left \{ \bar{P}_{v,b}^{\rm r}(r_0)>\bar{P}_{v,w}^{\rm r}(R_w) \right \},\\
\end{split}
\end{equation}
where $\xi_{v,b_1b_2}(r_0)$ gives the existence area of different types of BSs and is defined in (\ref{eq:rect}).
Based on Lemma~\ref{lemma2}, we obtain
\begin{equation}\label{eq:assoprob2}
\begin{split}
&\mathbb{P}\left \{ \bar{P}_{v,b}^{\rm r}(r_0)>\bar{P}_{v,w}^{\rm r}(R_w) \right \}=\mathbb{P}\left \{ R_w>r_{v,w|b}(r_0) \right \}\\
&\overset{(a)}{=} 1-F_{v,R_{w}}(r_{v,w|b}(r_0)) 
\overset{(b)}{=}\int_{r_{v,w|b}(r_0)}^{\infty } f_{v,R_w}(z){ \mathrm{d}z},
\end{split}
\end{equation}
where (a) follows the method in (\ref{eq:FUM}) and (b) follows the relationship between the CDF and the PDF. Substituting \eqref{eq:assoprob2} into \eqref{eq:assoprob1}, we complete the proof of Lemma~\ref{lemma3}.

\section{Proof of Lemma~\ref{lemma4}}\label{app:lemma4}
As defined in Sec.~\ref{subsec:channel},
$I_{v|r_0}$ is the interference experienced by the typical $v$-user conditioned on that the typical $v$-user is associated with the nearest BS located at $t_0$ with type $b$ and horizontal distance $r_0=||t_0||$. The Laplace transform of $I_{v|r_0}$ is given by
{
\begin{equation}\label{eq:L1}
\begin{split}
\mathcal{L}_{I_{v|r_0}}(s)
&=\mathbb{E}_{I_{v|r_0}}\left [ \exp (-sI_{v|r_0}) \right ]
\overset{(a)}{=}\mathbb{E}_{I_{v|r_0}}\left \{ \exp (-s\sum_{w\in W_v} I_{v,w|r_0}) \right \}\\ 
&=\mathbb{E}_{I_{v|r_0}}\left \{ \prod_{w\in W_v}\exp(-sI_{v,w|r_0}) \right \}
\overset{(b)}{=}\prod_{w\in W_v}\mathbb{E}_{I_{v,w|r_0}}
\left \{ \exp(-sI_{v,w|r_0}) \right \} ,
\end{split}
\end{equation}
where (a) is from \eqref{eq:interference}, $I_{v,w|r_0}$ is the interference from all interfering BSs in $\Psi_w$, and (b) follows the independence of interference from different types of BSs.
}
From (\ref{eq:powerrec}), (\ref{eq:interference}) and (\ref{eq:L1}), we have
\begin{equation}\label{eq:L2}
\begin{split}
&\mathbb{E}_{I_{v,w|r_0}} \left \{ \exp(-sI_{v,w|r_0}) \right \} 
=\mathbb{E}_{I_{v,w|r_0}} \left \{ \exp(-s\sum_{i,t_i\in \Psi_w\setminus \left \{ t_0 \right \} }P_{v,w}^{\rm r}(r_i)) \right \} \\
&=\mathbb{E}_{\Psi_w,\left \{ \Omega_{w_3,i} \right \}} \left \{ \prod_{i,t_i\in \Psi_w \setminus \left \{ t_0 \right \} }\exp(-sP^{\rm t}G_{w_2}\zeta_{v,w_3}\left ( r_i \right ) \Omega_{w_3,i}) \right \} \\
&\overset{(a)}{=}\mathbb{E}_{\Psi_w}\left \{ \prod_{i,t_i\in \Psi_w \setminus \left \{ t_0 \right \} } \mathbb{E}_{\Omega_{w_3}} \left \{ \exp (-sP^{\rm t}G_{w_2}\zeta_{v,w_3}(r_i) \Omega_{w_3}) \right \} \right \} \\
&\overset{(b)}{=}  \mathbb{E}_{\Psi_w}\left \{ \prod_{i,t_i\in \Psi_w \setminus \left \{ t_0 \right \} } \kappa_{w}(r_i,s) \right \}, \\
\end{split}
\end{equation}
where $\kappa_{w}(r_i,s)$ is given in (\ref{eq:L}), (a) follows the independence of PPPs and small-scale fading, and (b) follows the Gamma distribution of $\Omega_{w_3}$.
Employing the probability generating functional (PGFL) of PPP in~\cite{andrews2016primer}, (\ref{eq:L2}) can be further expressed as
\begin{equation}\label{eq:L3}
\begin{split}
\mathbb{E}_{I_{v,w|r_0}} \left \{ \exp(-sI_{v,w|r_0}) \right \} 
\overset{(a)}{=}\exp (-2\pi\lambda_{w_1}\int\limits_{r_{v,w|b}(r_0)}^{\infty} [1-\kappa_{w}(z,s)]z\mathcal{P}_v^{\rm w_3}(z)\xi_{v,w_1w_2}(r_0)\mathrm{d}z),
\end{split}
\end{equation}
where (a) is from the restriction of the nearest interfering BS in each type of BSs in Lemma~\ref{lemma2} and $r_{v,w|b}(r_0)$ is given in \eqref{eq:rwb}.
Substituting (\ref{eq:L3}) into (\ref{eq:L1}), Lemma~\ref{lemma4} has been proofed.

\section{Proof of Lemma~\ref{lemma5}}\label{app:lemma5}
Since we have already derived the expression for association probability $\mathcal{A}_{v,b}(r_0)$ in (\ref{eq:assoprob}) and distance distributions $f_{v,R_b}(r_0)$ in (\ref{eq:fUML})-(\ref{eq:fDSN}), the coverage probability in (\ref{eq:covprob0}) and (\ref{eq:covprob1}) can be further expressed as
\begin{equation}\label{eq:covprob3}
\begin{split}
\mathcal{P}^{\rm cov}_v
&{=\sum_{b\in W_v} \mathbb{E}_{R_b}\left [ \mathbb{P}\left \{ \mathcal{C}_{v}|\mathcal{B}_{v,b} ,R_b\right \}\mathbb{P}\left \{\mathcal{B}_{v,b}|R_b \right \}
 \right ]}
=\sum_{b\in W_v}\mathbb{E}_{R_b} \left \{ \mathcal{P}^{\rm cov}_{v,b}(R_b)\mathcal{A}_{v,b}(R_b)| R_b=r_0\right \}\\
&=\sum_{b\in W_v}\int_{0}^{\infty} \mathcal{P}^{\rm cov}_{v,b}(r_0)\mathcal{A}_{v,b}(r_0)f_{v,R_b}(r_0)\mathrm{d}r_0 ,
\end{split}
\end{equation}
where $\mathcal{P}^{\rm cov}_{v,b}(r_0)$ denotes the probability of the event that the typical $v$-user served by a $b$-BSs is in coverage conditioned on the serving distance $r_0$, i.e., $\mathcal{P}^{\rm cov}_{v,b}(r_0)={ \mathbb{P}\left \{ \mathcal{C}_{v}|\mathcal{B}_{v,b},r_0 \right \}}$,
and has a similar definition of $\mathcal{P}^{\rm cov}_v$ in (\ref{eq:covprob0}). From (\ref{eq:powerrec}), (\ref{eq:interference}) and (\ref{eq:sinrvb}), we have
{
\begin{equation}\label{eq:proof_pcov1}
\begin{split}
\mathcal{P}^{\rm cov}_{v,b}(r_0) 
&=\mathbb{P}\left \{ {\rm SIR}_v^b>\tau\mid R_b=r_0 \right \}
=\mathbb{P}\left \{ \frac{P_{v,b}^{\rm r}(r_0)}{I_{v \mid r_0}}> \tau \right \}\\
&=\mathbb{P}\left \{ \frac{P^{\rm t}G_{b_2}\zeta_{v,b_3}\left ( r_0 \right ) \Omega_{b_3,0}}{I_{v \mid r_0}}> \tau \right \}
=\mathbb{P}\left \{ \Omega_{b_3,0}> \frac{\tau I_{v \mid r_0}}{P^{\rm t}G_{b_2}\zeta_{v,b_3}\left ( r_0 \right )} \right \},\\
\end{split}
\end{equation}
where $I_{v|r_0}$ is given in (\ref{eq:interference}). }
Since the Gamma distribution of $\Omega_{b_3,0}$, (\ref{eq:proof_pcov1}) can be transformed into
{
\begin{equation}\label{eq:proof_pcov2}
\begin{split}
&\mathcal{P}^{\rm cov}_{v,b}(r_0) 
=\mathbb{E}_{I_{v|r_0}}\left \{ \mathbb{P}\left \{ \Omega_{b_3,0}> \frac{\tau I_{v|r_0}}{P^{\rm t}G_{b_2}\zeta_{v,b_3}\left ( r_0 \right )} \mid I_{v|r_0}\right \} \right \} 
{=}\mathbb{E}_{I_{v|r_0}}\left \{ \frac{\Gamma_u\left ( m_{b_3},sI_{v|r_0} \right )}{\Gamma\left ( m_{b_3} \right ) }   \right \} \\
&\overset{(a)}{=} \mathbb{E}_{I_{v|r_0}}\left \{ \sum_{k=0}^{m_{b_3}-1} \frac{s^k}{k!}{I_{v|r_0}}^{k} \exp\left ( -sI_{v|r_0} \right )\right \} 
=\sum_{k=0}^{m_{b_3}-1} \frac{s^k}{k!}\mathbb{E}_{I_{v|r_0}}\left \{ {I_{v|r_0}}^{k} \exp\left ( -sI_{v|r_0} \right ) \right \} ,
\end{split}
\end{equation}
}where $\Gamma_u\left ( m,mg \right )=\int_{mg}^{\infty} t^{m-1}e^{-t}\mathrm{d}t$, $s$ is given in (\ref{eq:covprob_vb}),
and (a) is from the definition $\frac{\Gamma_u\left ( m,g \right )}{\Gamma\left ( m \right )} =\exp(-g) {\textstyle \sum_{k=0}^{m-1}}\frac{g^k}{k!}$.
Applying the property of Laplace transform into (\ref{eq:proof_pcov2}), we have
\begin{equation}\label{eq:proof_pcov3}
\begin{split}
\mathcal{P}^{\rm cov}_{v,b}(r_0) 
= \sum_{k=0}^{m_{b_3}-1} \frac{s^k}{k!}\left ( -1 \right )^k \frac{\mathrm{d^k} }{\mathrm{d^k} s} \mathcal{L}_{I_{v|r_0}}(s)
=\sum_{k=0}^{m_{b_3}-1} \frac{(-s)^k}{k!}\frac{\mathrm{d^k} }{\mathrm{d^k} s} \mathcal{L}_{I_{v|r_0}}(s)
\end{split}.
\end{equation}
With \eqref{eq:proof_pcov2} and \eqref{eq:proof_pcov3}, we obtain the final expression of $\mathcal{P}^{\rm cov}_v$ in Lemma~\ref{lemma5}.


\section{Proof of Lemma~\ref{lemma6}}\label{app:lemma6}
The CDF of Gamma distribution is $F_G(g)=\frac{\Gamma_l\left ( m,mg\right )}{\Gamma\left ( m \right ) }$, where  $\Gamma_l\left ( m,mg \right )=\int_{0}^{mg} t^{m-1}e^{-t}\mathrm{d}t$.
The authors in~\cite{gamma} give the upper bound and the lower bound of $F_G(g)$, which is given by
\begin{equation}\label{eq:gamma_bound}
\left [ 1-\exp\left ( 1-\beta_1 mg  \right ) \right ]  ^{m} <\frac{\Gamma_l\left ( m,mg\right )}{\Gamma\left ( m \right ) }< \left [ 1-\exp\left ( 1-\beta_2 mg  \right ) \right ]  ^{m},
\end{equation}
where $m\ne 1$ and 
\begin{equation}\label{eq:gamma_beta}
\beta_1=\left\{\begin{matrix}
 1\hfill& {\rm if}~ m>1\\
 (m!)^{\left ( \frac{-1}{m}  \right ) } & {\rm if}~m<1,
\end{matrix}\right.
~\beta_2=\left\{\begin{matrix}
 (m!)^{\left ( \frac{-1}{m}  \right ) } &{\rm if}~ m>1\\
 1\hfill  &{\rm if}~m<1.
\end{matrix}\right.
\end{equation}
Using the upper bound in (\ref{eq:gamma_bound}), we rewrite (\ref{eq:proof_pcov2}) as 
{
\begin{equation}\label{eq:approximate1}
\begin{split}
&{\mathcal{P}} ^{\rm cov}_{v,b}(r_0)
=\mathbb{E}_{I_{v|r_0}}\left \{ \frac{\Gamma_u\left ( m_{b_3},s I_{v|r_0} \right )}{\Gamma\left ( m_{b_3} \right ) }   \right \}
=1-\mathbb{E}_{I_{v|r_0}}\left \{ \frac{\Gamma_l\left ( m_{b_3},s I_{v|r_0} \right )}{\Gamma\left ( m_{b_3} \right ) }   \right \}\\
&\overset{(a)}{\approx}  1-\mathbb{E}_{I_{v|r_0}}\left \{ \left ( 1-e^{-\beta_{b_3}sI_{v|r_0}} \right )^{m_{b_3}} \right \} 
=\sum_{k=1}^{m_{b_3}}\binom{m_{b_3}}{k}\left ( -1 \right )^k \mathcal{L}_{I_{v|r_0}}\left ( k \beta_{b_3} s \right )   
\overset{\triangle}{=}\tilde{\mathcal{P}} ^{\rm cov}_{v,b}(r_0),
\end{split}
\end{equation}
}where $\Gamma_l\left ( m,mg \right )=1-\Gamma_u\left ( m,mg \right )$ and (a) follows the method proofed in~\cite{approximate}.
The approximate form of $\mathcal{P} ^{\rm cov}_{v,b}(r_0)$ in (\ref{eq:approximate1}) simplifies (\ref{eq:covprob1}) into
\begin{equation}\label{eq:approximate2}
\begin{split}
\tilde{\mathcal{P}} ^{\rm cov}_v
=\sum_{b\in W_v}\int_{0}^{\infty} \tilde{\mathcal{P}} ^{\rm cov}_{v,b}(r_0)\mathcal{A}_{v,b}(r_0)f_{v,R_b}(r_0)\mathrm{d}r_0.
\end{split}
\end{equation}
Finally, $\tilde{\mathcal{P}} ^{\rm cov}_v$ is derived after substituting (\ref{eq:approximate1}) into (\ref{eq:approximate2}).



\bibliographystyle{IEEEtran}
\bibliography{reference}
\end{document}